\documentclass[referee,usenatbib]{mn2e}
\usepackage{epsfig}
\usepackage{subfigure}
\usepackage{cite}
\newcommand{\msun}{\>{\rm M_{\odot}}}

\newcommand{\beq}{\begin{equation}}
\newcommand{\eeq}{\end{equation}}

\newcommand{\Obaryon}{{\Omega_{\rm B,0}}}

\newcommand{\msunh}{\>h^{-1}\rm M_\odot}

\newcommand{\calT}{{\cal T}}

\newcommand{\mvir}{m_{\rm vir}}

\newdimen\hssize
\newdimen\hdsize 
\begin{document}

\title[Bayesian inference from the $K$-band luminosity function]
	{Bayesian inference of galaxy formation from the $K$-band luminosity function
	of galaxies: tensions between theory and observation}
\author[]
       {Yu Lu$^{1}$\thanks{E-mail: luyu@stanford.edu}, H.J. Mo$^{2}$,
         Neal Katz$^{2}$, Martin D. Weinberg$^{2}$
\\
		$^1$ Kavli Institute for Particle Astrophysics and Cosmology, 
		Stanford, CA 94309, USA
\\
        $^2$ Department of Astronomy, University of Massachusetts,
        Amherst MA 01003-9305, USA
		}


\date{}

\pagerange{\pageref{firstpage}--\pageref{lastpage}}
\pubyear{2012}

\maketitle

\label{firstpage}


\begin{abstract}
  We conduct Bayesian model inferences from the observed $K$-band
  luminosity function of galaxies in the local Universe, using the
  semi-analytic model (SAM) of galaxy formation introduced in
  \citet{Lu2011a}.  The prior distributions for the 14 free
    parameters include a large range of possible models.  We find
  that some of the free parameters, e.g. the characteristic scales for
  quenching star formation in both high-mass and low-mass halos, are
  already tightly constrained by the single data set.  The
  posterior distribution includes the model parameters adopted
  in other SAMs.  By marginalising over the posterior distribution, we
  make predictions that include the full inferential uncertainties for
  the colour-magnitude relation, the Tully-Fisher relation, the
  conditional stellar mass function of galaxies in halos of different
  masses, the HI mass function, the redshift evolution of the stellar
  mass function of galaxies, and the global star formation history.
  Using posterior predictive checking with the available observational
  results, we find that the model family (i) predicts a Tully-Fisher
  relation that is curved; (ii) significantly over predicts the
  satellite fraction; (iii) vastly over predicts the HI mass function;
  (iv) predicts high-$z$ stellar mass functions that have too many
  low mass galaxies and too few high mass ones.
  and (v) predicts a redshift evolution of the stellar mass density
  and the star formation history that are in moderate disagreement.
  These results suggest that some important processes are still
  missing in the current model family and we discuss a number of
  possible solutions to solve the discrepancies, such as interactions
  between galaxies and dark matter halos, tidal stripping, the bimodal
  accretion of gas, preheating, and a redshift-dependent initial mass
  function.
\end{abstract}



\section{Introduction}\label{sec:introduction}

During the past 20 years semi-analytic models of galaxy formation have
been developed and widely adopted to study the statistical properties
of the galaxy population in the cold dark matter (CDM) cosmogony
\citep[e.g.][]{White1991, Lacey1991, Kauffmann1993, Cole1994, Mo1998,
  Somerville1999,Cole2000, Kang2005, Croton2006, Dutton2009}.  In a
semi-analytic model (hereafter SAM), one adopts recipes to describe
and parametrise the underlying physical ingredients, such as star
formation and feedback. The free parameters in the models are then
tuned to reproduce certain observational properties of the galaxy
population.  Since a variety of the physical processes that affect
galaxy formation and evolution are still poorly understood
\citep[e.g][]{Mo2010}, one must quantitatively characterise the model
constraints implied by the existing data sets as well as explore a
wide range of models. The SAM approach provides a promising avenue to
fulfil these tasks owing to its flexibility in implementation and its
relatively fast speed in computation. By translating the theory of
galaxy formation into a set of model parameters, SAMs can be used to
make model inferences from observational data, and to make predictions
for further tests of the theory.

In conventional implementations of SAMs, model inferences and
predictions are performed in two steps.  One first
tunes the model against a set of observational
constraints to find a parameter set, and then uses
this parameter set to make predictions for other observables.
For example, one can choose model parameters governing
the efficiency of star formation and stellar energy feedback so that a
``Milky Way'' halo contains, on average, the same mass in stars and
cold gas as our own Galaxy, and adjust the dynamical friction time
scale so that a ``Milky Way'' halo contains, on average, the right
number of ``Magellanic Cloud''-sized satellites. Such requirements,
together with the Tully-Fisher relation, have been used by
\citet[][]{Kauffmann1999, Somerville1999, DeLucia2004} to tune
their SAMs. Alternatively, \citet[][]{Cole1994, Cole2000, Kang2005,
  Baugh2005, Bower2006}, among others, chose to tune their SAMs
with the observed luminosity functions of galaxies.

There are a number of problems with such implementations.  First,
a satisfactory fit is typically assessed by eye so that
uncertainties in the observational data, a crucial aspect in model
inference, is not properly taken into account.  Second, the
tuning is usually done by ``hand''.  One typically adjusts
one parameter at a time until one obtains a satisfactory fit to
the data. The problem is that the likelihood function is typically
complex and it is very difficult to find the global optimum with hand
tuning \citep{Lu2011a}.  Third, this approach only yields a
single parameter set so that model predictions are made without
including uncertainties in the inference.  Such predictions are
questionable because of the fact that the model parameters are largely
degenerate \citep{Bower2010, Lu2011a, Neistein2010}.  For all the
above reasons, the full potential power of SAMs has not been fully
realised. To derive meaningful constraints from observations and to
make reliable predictions, one needs to know the relative probability
of various model parameters and, indeed, the probability of entire
model families given some set of observational data. This is best
achieved by a Bayesian inference.

In \citet{Lu2011a}, we have developed a scheme to incorporate SAMs
into the framework of Bayesian inference. To this end, we have
constructed a general SAM that contains a number of published SAMs as
subsets. We have also shown that, aided with advanced MCMC techniques
and parallel computation, it is now
possible to build a Bayesian inference-based SAM to efficiently
explore the high dimensional parameter space and to establish the
posterior distribution of the model parameters reliably. In the Bayesian
framework, one first chooses a set of observations as data constraints
to derive the posterior probability distribution of the model
parameters.  Such a posterior distribution encompasses the
uncertainties of the model parameters and the
  observational constraints, allowing one to assess the theory in a
statistically rigorous way. Furthermore, model predictions can be made
from the posterior distribution including these
uncertainties, providing an avenue to assess the power of future
observations.  The general scheme of the method along with some simple
examples have been presented in
\citet{Lu2011a}.  Following our previous paper, we
review Bayesian model
prediction and introduce a systematic model checking
procedure  called Bayesian posterior
predictive checking \citep[PPC,][]{Gelman2004, Gilks1995}.
This technique exploits the full predictive power of SAMs.

In this paper, we start to put the conventionally, though potentially flawed, 
two-step procedure of first fixing the model parameters using one data set and
then using them to predict other observations on a firm statistical footing. 
We apply Bayesian inference with PPC 
to the observed $K$-band luminosity function of local
galaxies, taking into account all the known uncertainties in the data,
to derive realistic constraints on the model parameters of an extended
model family.  First, we use observations to constrain a SAM
and explore the implications of these constraints on the underlying
physical processes affecting galaxy formation and evolution.  Second,
we use the full posterior distribution of the model parameters
obtained to make predictions for a number of other observable
properties of galaxies. These include (i) the cold gas mass function
of galaxies; (ii) the Tully-Fisher relation; (iii) the
colour-magnitude relation of galaxies; (iv) the conditional stellar
mass function of galaxies in halos of various masses; and (v) the
redshift evolution of the stellar mass function of galaxies, the star
formation rate density, and the cold gas mass density.  These model
predictions are compared with available observational data to check
whether the current model family is able to accommodate the
observational results.  For most of the predictions, we apply the
quantitative Bayesian model checking method introduced
here to assess the model.  For
other predictions, for example the cosmic stellar mass density as a
function of redshift, the cosmic star formation rate density as a
function of redshift, and the cold gas mass density as a function of
redshift, the error model for the comparison is uncertain. For these 
predictions, we only check the model with the available data graphically.

The paper is organised as follows.  In \S\ref{sec:bayes}, we describe
our Bayesian approach, including the MCMC technique for sampling the
posterior, Bayesian model prediction, and Bayesian posterior
model checking.  \S\ref{sec:model} briefly describes our
semi-analytic model and \S\ref{sec:data} presents the data used to
constrain the model and defines the likelihood function.
In \S\ref{sec:constraint}
we show how the $K$-band luminosity function of galaxies constrains
the posterior distribution of the model parameters.  In
\S\ref{sec:prediction}, we use the posterior obtained to make
predictions for the colour-magnitude relation, the Tully-Fisher
relation, the cold gas mass function, the conditional stellar mass
function, and the redshift evolution of the stellar mass function, and
compare them with observational data.  Finally, we summarise and discuss our
results further in \S\ref{sec:discussion}.

Throughout the paper, we use a $\Lambda$CDM cosmology
with $\Omega_{\rm M,0}=0.26$, $\Omega_{\rm \Lambda,0}=0.74$,
$\Obaryon=0.044$, $h=0.71$, $n=0.96$, and $\sigma_8=0.80$.
These values are consistent with the WMAP5
data \citep{Dunkley2009, Komatsu2009}.

\section{The Bayesian Approach}
\label{sec:bayes}

\subsection{MCMC simulations of the posterior distribution}
\label{sec:bayes_mcmc}

As detailed in \citet{Lu2011a}, a variety of physical processes
affecting galaxy formation are not yet well understood while copious
observational data constrain the models. To derive meaningful
constraints from the observations, we need to know the probability of
the model parameters given the data. The Bayesian approach allows us
to obtain this posterior distribution of the
model parameters for a given set of data and to make robust
predictions taking into account uncertainties present in the model.
To sample the posterior probability distribution, we employ the
Bayesian Inference Engine \citep[BIE,][]{Weinberg2010}
\footnote{\tt http://www.astro.umass.edu/BIE}, which
includes a suite of advanced Markov-Chain Monte-Carlo (MCMC)
algorithms and supports parallel computation.  In particular, we adopt
the Tempered Differential Evolution (TDE) algorithm to sample the
posterior. The MCMC algorithm provides proposal parameter vectors for
the SAM, and the SAM predicts the galaxy population for the given set
of model parameters. The likelihood of the data
given the model is evaluated and returned to the MCMC
program.  The MCMC algorithm accepts or rejects the proposal based on
the posterior probability, and generates a new proposal for the
SAM. To ensure that the chains have sufficiently explored the
parameter space, we first run the MCMC at a
higher temperature, namely we sample the more diffuse distribution function
defined by
\begin{equation}
p'(\theta|D) \propto p(\theta|D)^{1/T_{\rm p}}, 
\end{equation}
where $p(\theta|D)$ is the real posterior probability, and $T_{\rm p}$
is the so-called powered-up temperature
with $T_{\rm p}\ge1$. Since $p^\prime(\theta|D)$
is more diffuse (i.e. flatter) than $p(\theta|D)$, the Markov chain
can jump out of a local mode with higher probability
and hence explore a larger range of parameter space.  After the chains
are converged using the hotter state, we resume the simulation from the
current states at the fiducial temperature, i.e. $T_{\rm p}=1$.  The
MCMC simulation again continues until convergence is achieved. The
convergence of the chains is monitored by the Gelman-Rubin ${\hat R}$
statistic \citep{Gelman1992}, and we declare convergence when ${\hat R}\leq 1.2$.

\subsection{The posterior predictive distribution}
\label{sec:bayes_predition}

Once we have the posterior distribution, we can make predictions 
for other observables by marginalising the desired
  likelihood function over the posterior 
\citep[e.g][]{Gelman2004}.
The predicted distribution of a new observable, $\mathbf{y'}$,
given the data constraint $\mathbf{y_c}$, is
\begin{equation}
p({\mathbf {y'}}|{\mathbf{y_c}})=\int p({\mathbf{y'}}|{\mathbf{\theta}}) 
p({\mathbf{\theta}}|{\mathbf{y_c}}) {\rm d}{\mathbf{\theta}}\,,
\label{eq-ppd}
\end{equation}
where $\mathbf{\theta}$ denotes the model parameter vector,
$p({\mathbf{\theta}}|{\mathbf{y_c}})$ is the posterior distribution
obtained from the data constraint $\mathbf{y_c}$, and
$p({\bf{y'}}|{\mathbf{\theta}})$ is the probability distribution of
the observable ${\bf{y'}}$ for a given model specified by
$\mathbf{\theta}$ (i.e. a likelihood function).  For
deterministic models, the distribution function of predicted
observations, $p({\bf{y'}}|{\mathbf{\theta}})$ is a $\delta$ function,
$\delta[{\bf{y'}}-{\bf{y'}}({\mathbf{\theta}})]$.  For probabilistic
models, if the variance of the prediction $\bf{y'}$ from a given model
$\mathbf{\theta}$ is much smaller than the variance from the
posterior, the $\delta$ function is also a good approximation.  The
resulting distribution function $p({\bf{y'}}|{\mathbf{y_c}})$, called
the posterior predictive distribution (PPD), encompasses all the
inferential uncertainties and hence provides the confidence level of
the predicted observable. For a complex model like the SAM considered
here, the PPD can be obtained using MCMC samples. To do this, one
first selects a sample from the converged posterior distribution $\{\mathbf{\theta\}}$.  
For each of the $\mathbf{\theta}\in\{\mathbf{\theta}\}$ selected, the predictions of
$\bf{y'}$ are obtained from the probability distribution,
$p(\bf{y'}|\mathbf{\theta})$, and these $\bf{y'}$
are a sample of the PPD.

\subsection{The posterior predictive check}
\label{sec:bayes_check}

Once the posterior predictive distribution is obtained, one can check
the specific model family using a procedure called posterior
predictive check \citep[hereafter PPC,][]{Gelman2004, Gilks1995}. The
central idea of PPC is that the data replicated 
from the model should be distributed as the observed data.  
Any discrepancies then indicate that the
model may be incorrectly specified.  The PPC also applies to new observables
that are not included in the data constraint. If the model is true,
the PPD should not show a large inconsistency with the data of those
predicted observables.

In practise, a graphical representation of the PPD and
the data distribution or its summary quantities may be sufficient to identify
discrepancies. The later
option is particularly useful when the data set is large or a
particular aspect of the data contains important information.  When
graphical PPCs do not reveal the
discrepancies between the models and data, 
one can perform numerical PPCs, which are a quantitative
measure of the discrepancies. To perform a numerical PPC, one first
needs to define a test statistic ${\calT}({\bf{y}}, {\mathbf{\theta}})$ 
designed to discriminate between the model and the data. 
In general, the test statistic may depend
explicitly on ${\mathbf \theta}$, but in our applications such a
dependence is absent and so in the following we will omit ${\mathbf
  \theta}$ from the independent variables of ${\calT}$.  In this
paper, we use the tail-area probability (e.g. the $p$-value) of the
test statistic $\calT({\bf y})$ to assess the lack of a fit to the
data.

To motivate a {\it Bayesian} definition of a $p$-value test, 
we first note that the {\it classical} $p$-value is defined as
\begin{equation}
p_C =P\left[\calT({\bf y'}) \geq \calT({\bf y})|{\mathbf{\theta}}\right]\,,	
\end{equation}
where the probability, $P$, is calculated over the distribution of 
${\bf y'}$ with ${\mathbf\theta}$ fixed. 
In classical testing, ${\mathbf\theta}$ would correspond to 
the null hypothesis value. It could also be a point estimate such 
as a maximum likelihood estimate. In the Bayesian context, we 
can generalise the test statistic to allow for a dependence on the 
model parameters under their posterior distribution, so that
both the variance of the observations ($\bf{y}$) and the 
uncertainties of the parameter values ($\mathbf{\theta}$) are 
taken into account. Thus the {\it Bayesian} $p$-value is
\begin{equation}\label{eq-pB}
p_B	=P\left[\calT({{\bf y'}})\geq \calT({\bf{y}})  | {{\bf y_c}}\right] 
        = \int \int I_{\calT ({\bf{y'}}) \geq\calT({\bf{y}})} 
         \,p({\mathbf{y'}} |{\mathbf{\theta}})\,p({\mathbf{\theta}}|{\bf{y_c}}) 
         \rm{d} {\bf{y'}} {\rm d} {\mathbf \theta},
\end{equation}
where $I_q$ is the indication function for the condition $q$ ($I_q$
equals $1$ if $q$ is true and $0$ otherwise). Note that the testing data 
${\bf y}$ can be the same as the constraining data ${\bf y_c}$ or some other data. 
If the predicted observables $\bf{y'}$ are incompatible with the model,
then the observed test statistic $\calT({\bf y})$
may be a significant outlier of the distribution of
the test statistic $\calT({\bf{y'}})$ predicted by the model.  If the
posterior predictive $p$-value is close to 0 or 1 (typically chosen to
be 0.05 or 0.95), then the model is
most likely inadequate. Note that this approach is
similar to classical hypothesis testing, where a test statistic
$\calT$ measures the discrepancy between the data and the predictive
simulations.

Usually we cannot calculate the Bayesian $p$-value analytically, but we
can do it using posterior simulations. 
Suppose that we have $L$ samples of ${\mathbf{\theta}}$, 
$(\theta_1, \cdot\cdot\cdot , \theta_L)$, randomly drawn 
from the posterior distribution $p({\mathbf \theta}|{\bf y_c})$. 
Then for each of these ${\mathbf{\theta}}$ samples, we can 
generate one sample ${\bf y'_l}$ from $p({\mathbf{y'}}|{{\mathbf \theta}}_l)$. 
The Monte Carlo evaluation of equation (\ref{eq-pB}) is then
\begin{equation}
\hat{p}_B = {1 \over L} \sum_{l=1}^L I_{\calT({\bf y}'_l)
  \geq\calT({\bf y})}\,,
\end{equation}
where ${\bf y}'_l$ is the prediction of sample ${\bf\theta}_l$.
In other words, the fraction of samples where $\calT({\bf y}') \geq
\calT({\bf y})$ is an estimate of $p_B$. 
Note that the test statistic $\calT (\bf{y})$ needs to be chosen to
effectively investigate the deviations of interest. 
This is similar to choosing a powerful test statistic when conducting 
a hypothesis test.

The Bayesian PPC includes all the inferential
uncertainties implied by the
constraining data and provides confidence
bounds for the predicted quantities. Any significant inconsistency
between the predictions and the data 
suggests that modifications to the model are required. As such, the
Bayesian PPC provides a powerful method to test the admissibility of models
given data.  The method, however, also has its
limitations because PPC does not provide a probability for rejecting a model.
First, if a model family passes a PPC, it does not
necessarily mean the model family is free of problems; it may only
have passed because the chosen test statistic was
insufficiently powerful. Therefore, the choice of the test statistics is
crucial.
Second, although a large difference between the PPD and the
observational data indicates tensions between the model and the data,
one cannot reject a model family, because an improper prior distribution 
can also result in a biased PPD.
The only way to
identify modes that can simultaneously explain multiple data sets is
to perform the Bayesian inference using the full data sets, and to use
the posterior to conduct a Bayesian goodness-of-fit test.

In this paper, we will use PPC to identify tensions between our SAM
and a variety of existing data sets as follows.  Suppose that
we have drawn $L$ samples from the posterior distribution 
and that the predicted observables are
given in $N$ bins.  Denote the value
of the prediction of the $l$th parameter vector
in the $i$th bin by $y'_{l,i}$. We define
\begin{equation}\label{eq-calTmodel}
\calT_l\equiv \calT({\bf y}'_l)
=\sum_{i=1}^N {\left( y'_{l,i}- {\overline y'}_i\right)^2\over
 \sigma'^2_i}\,,
\end{equation}
where ${\overline y}'_i$ and $\sigma'^2_i$ are, respectively, the mean
and the standard deviation obtained from the $L$ posterior samples\footnote{The
model is specified, and it is the parameters that are being
sampled.}.   The histograms of $\calT_l$
($l=1, \cdot\cdot\cdot, L$) are then used to represent the probability
distribution of the test statistic $\calT$ predicted by the model.  To
compare with the observations, we define a similar test statistic from
the observational data $y_i$:
\begin{equation}\label{eq-calTdata}
\calT^{\rm obs}\equiv \calT({\bf y})
=\sum_{i=1}^N {\left(y_i- {\overline y}'_i\right)^2\over
 \sigma'^2_i}\,, 
\end{equation}
where ${\overline y}'_i$ and 
$\sigma'^2_i$ are the same as in equation (\ref{eq-calTmodel}). 
Comparing the test quantity from the observations with the distribution 
of the test quantity predicted by the model, the $p$-value then tells
us the odds of having such observational data given the constrained 
model.

If there are $M$ independent observations of ${\bf y}_m$ ($m=1,\cdot\cdot\cdot,M$), 
then following equation (\ref{eq-calTmodel}) one can compute
the test quantity for each of the $M$ observations:
\begin{equation}
\calT_m^{\rm obs}\equiv \calT({\bf y}_m)
=\sum_{i=1}^N {\left(y_{m,i}- {\overline y'}_i\right)^2\over
\sigma'^2_i}\,. 
\end{equation}
The histograms given by $\calT_m^{\rm obs}$ 
($m=1, \cdot\cdot\cdot, M$) then represent the 
probability distribution of  $\calT^{\rm obs}$. Data are often 
presented as the mean of a quantity together with
error bars that describe the uncertainties of the measurements. 
In this case, one can generate $M$ replica according to 
the error budget in the data and construct the distribution of 
the test quantity from these replica to take into account the
observational uncertainties. The difference between the model 
and the data is then given by comparing the distribution
of $\calT_l$ and that of  $\calT^{\rm obs}_m$.  
In our following applications, we treat the observational mean as one
realisation of the observable in question. In this case, we calculate the
value of $\calT^{\rm obs}$ from equation (\ref{eq-calTdata}), now with
${\bf y}$ set to be the observational mean, and compare it with the
distribution of $\calT_l$.

One potential problem of a test based on a $\chi^2$-like test
quantity, like that described above, is that the power of the test may be
diminished when the bins are strongly correlated and
the number of bins is large.  The resulting relatively large number
of dependent variables will weaken the power of $\calT$.  The power can be 
improved by incorporating the covariance and including only 
the independent degrees of freedom. Principal component
analysis (PCA) can achieve this using the following
widely-used procedure \citep{Murtagh1987}. We first construct a data
matrix ${\bf Y}'$ from $L$ model predictions of ${\bf y}'_l$
($l=1,\cdot\cdot\cdot,L$), which has $N$ columns and $L$ rows.  We
then zero the centre by the mean, $\overline{\bf y}'_i$, and scale the data
by the standard deviation, $\sigma_{i}$, of each column of the
data matrix, yielding ${\bf Y_s}$. The PCA, which we perform using
singular value decomposition (SVD), yields $N$ unit eigenvectors,
${\bf e}_i$, and $N$ corresponding eigenvalues, $\lambda_i$.
We construct a $N \times N$ transformation matrix, ${\bf U}$, by
putting each eigenvector on each row.  In the matrix, the
eigenvectors are ordered so that the one with a larger eigenvalue is
put on a upper row.
The matrix of eigenvectors is a unitary transformation
of the data ${\bf Y_s}$ to a space where each dimension is
uncorrelated. The eigenvalues describe the variance of the data in
this new space. 
Using the transformation matrix, we find the transformed data transposed as 
${\bf X}'^{\mathrm T}= {\bf U} {\bf Y_s}^{\mathrm T}$. 
We may now write a new correlation-free test statistic for each row vector 
of the ${\bf X}'$ as:
  \begin{equation}
    \calT_l^{\prime} = \sum_{i=1}^R\frac{x^{\prime 2}_{l\,i}}{\lambda_i}.
    \label{eq-transT}
\end{equation}
where $R\le N$ will be specified below.  Equation (\ref{eq-transT}) is
not equivalent to equation (\ref{eq-calTmodel}) although it does have
a similar interpretation: we expect the $\calT_l^{\prime}$ to be
distributed as a multivariate normal distribution, appealing to the
central limit theorem in the large $L$ limit.  Assume that we have
ordered the eigenvalues such that $\lambda_i>\lambda_j$ if $i<j$.
Recall that the eigenvector
with the highest eigenvalue is the principal component of the data set
that preserves the largest uncorrelated fraction of the total initial variance.  
In many cases,
the magnitude of the $\lambda_j$ decreases quickly with increasing $j$.
Components with $\lambda_j\ll\lambda_1$ carry little information.
This allows us to truncate the summation in equation (\ref{eq-transT}) by
choosing $R$ to be the smallest integer such that $\sum_{i=R+1}^N
\lambda_i/\sum_{i=1}^N\lambda_i<\epsilon_R$. Here, we
choose ${\epsilon_R}=0.01$.  We find that this
criterion preserves most of the information of the posterior
prediction distribution and ignores the details that can not be
distinguished given the level of the observational errors present.
Once we determine the truncation component, we then define a
modified transformation matrix ${\bf U_R}$ by setting all the elements
except the first $R$ rows of the matrix ${\bf U}$ to zero, and compute
the transformation of the data as
${\bf X}'^{\mathrm T} = {\bf U_R} {\bf Y_s}^{\mathrm T}$. 
For the observational data vector, we follow exactly the same adjustment
and transformation defined by the prediction data matrix, which yields
the test quantity for the observations.  We use the reference
distribution of the test statistic (eq. \ref{eq-transT}) computed
from the posterior sample to calculate the $p$-value of the
observations.


\section{Model and Model Parameters}
\label{sec:model}

We employ the SAM developed by \citet{Lu2011a}, in which the
parameterisations for star formation and supernova (SN) 
feedback are generalised
to encompass many existing models.  Here we briefly describe the model
and readers can refer to \citet{Lu2011a} for more details.  Our SAM
starts with Monte Carlo derived halo merger trees \citep{Parkinson2008} using the current
$\Lambda$CDM model, and includes important physical processes for
galaxy formation, such as gas cooling, star formation, supernova (SN)
feedback, galaxy mergers, and AGN feedback. Gas is assumed to be
heated by accretion shocks and to form a hot gaseous halo that cools
by radiative cooling. Owing to a reduced cooling rate and 
heating by AGN feedback, gas cooling is assumed to be unimportant  
in massive halos. We model this using a free parameter, $M_{\rm CC}$, 
the halo mass above which radiative cooling becomes negligible. 
The cooling gas settles into the halo centre as a disk of cold gas, 
where stars form in regions where the surface density of the disk 
is sufficiently high. We  use a free parameter, $f_{\rm SF}$, to control 
the cold gas surface density threshold for star formation; only 
gas above the surface density threshold can form stars. 
The star formation efficiency is assumed to be proportional to 
the total cold gas mass for star formation and inversely proportional to the 
dynamical time scale of the disc, with a overall efficiency 
$\epsilon_\star$ assumed to be a broken power law of 
the halo circular velocity, $v_{\rm vir}$: $\epsilon_\star =\alpha_{\rm SF}$  
for $v_{\rm vir}>V_{\rm SF}$, and  
$\epsilon_\star =\alpha_{\rm SF}
(v_{\rm vir}/V_{\rm SF})^{\beta_{\rm SF}}$ 
for $v_{\rm vir}<V_{\rm SF}$.  
The amplitude, $\alpha_{\rm SF}$, the power index, $\beta_{\rm SF}$, 
as well as the pivotal circular velocity, $V_{\rm SF}$, are 
all treated as free parameters. 
The SN feedback associated with star formation is assumed to reheat a
fraction of the cold gas in the galaxy and may drive an outflow that
can remove some of the baryons from the host halo. 
The fraction of the total SN energy that affects subsequent 
star formation is assumed to be $\alpha_{\rm SN}$. 
The mass of the reheated cold gas is assumed 
to be proportional to the stellar mass, with the 
proportionality given by 
$f_{\rm rh}=\alpha_{\rm RH} (V_0/v_{\rm vir})^{\beta_{\rm RH}}$,  
where $V_0$ is set to be $220{\rm km\,s^{-1}}$, and 
$\alpha_{\rm RH}$ and $\beta_{\rm RH}$ are free parameters.   
A fraction of the SN energy that is not used to reheat 
the gas is assumed to drive galactic winds, and this 
fraction is controlled by a free parameter $\epsilon_{\rm W}$. 
Finally, a fraction of the ejected baryonic mass is assumed 
to come back to the halo as hot halo gas on a dynamical time scale, 
and this fraction is controlled by a free parameter, $f_{\rm RI}$.
When two or more dark matter halos merge, the central galaxy 
of the more massive halo is assumed to become the central 
galaxy of the new halo. The time over which a satellite 
galaxy orbits in its host halo before merging into the 
central galaxy is calculated based on the dynamical friction 
timescale of the secondary halo that hosts the satellite galaxy, 
and the real merging timescale is assumed to be a free 
parameter, $f_{\rm DF}$, times this dynamical friction 
timescale. When a satellite galaxy merges into a central
galaxy, a fraction of the total cold gas in these two merging 
galaxies is converted into stars through a starburst, and this
fraction is assumed to be given by the satellite-to-central  
mass ratio as $\alpha_{\rm SB} (m_{\rm sat}/m_{\rm cen})^{\beta_{\rm SB}}$, 
with $\alpha_{\rm SB}$ and $\beta_{\rm SB}$ two free parameters.   
A morphological transformation may occur depending on the mass 
ratio between the two merging progenitors. 

The processes included in our SAM are similar 
to those in other semi-analytic
models, but our parameterisations of the physical processes are
designed to cover many published SAMs as subsets and encompass
the physically plausible ranges for these processes.  
The free parameters are summarised in Table
\ref{tab:param}.  The prior ranges for the
parameters are listed in the last column of the table. The prior
distributions of the parameters (some are taken to be logarithmic) are
simply assumed to be uniform, as we have limited knowledge about them.
In total, we have 14 free parameters including a parameter describing
the incompleteness of the K-band luminosity function at the faint end
(see \S\ref{sec:data}).  We adopt a recently updated stellar
population synthesis model \citep{Bruzual2007} to convert the
predicted stellar masses into K-band light.

\section{Data and Likelihood}
\label{sec:data}

In \citet{Lu2011a} we adopted an approximate
stellar mass function of galaxies based on \citet{Bell2003}, instead
of the $K$-band luminosity function, as the observational constraint
just to demonstrate the viability of our Bayesian approach to SAMs of
galaxy formation and to illustrate some basic facts about the
approach.  However, as we showed in \citet{Lu2011a}, systematic errors
associated with measurements of the stellar mass function of galaxies
are difficult to treat in the likelihood function because their
statistical properties are not well understood, and an improper
treatment of the systematic errors can result in biased inferences.
In contrast, making a likelihood function in terms of the luminosity
function is rather straightforward because the luminosity of a galaxy
is a direct observable, and measuring the luminosity function is
simply a counting process. When systematic errors in the luminosity
measurement are negligible, the data in each luminosity bin is
independent. As our goal in the present paper is to derive a reliable
posterior, we choose to use the $K$-band luminosity function with a
realistic error model as our observational constraint.

The $K$-band luminosity function obtained by \citet{Bell2003} is based
on the 2MASS Extended Source Catalogue
\citep{Jarrett2000} with an incompleteness correction based on the
SDSS. To model the error budget, one may first
predict the properties of the galaxy population, and then simulate the
process of measuring the luminosity
function. Because the luminosity function could be incomplete for
faint galaxies, corrections for observational selection effects should
be made to the model prediction.  In what follows we introduce our
treatments for both counting errors and sample incompleteness.

We first formulate the counting process for the binned luminosity function.
When systematic errors in the luminosity measurements are negligible,
this is a Poisson process and we use the number
counts in each luminosity bin to compute the likelihood. Unfortunately,
the observational number counts are not available to us. Here we use
an alternative approach using the observed luminosity function to
obtain the number counts. For a given absolute magnitude, $M_i$, we
estimate the largest luminosity distance within which a galaxy with
such an absolute magnitude can be observed in a survey with an
apparent magnitude limit $m_{\rm lim}$.  The apparent magnitude limit
of the sample used in \citet{Bell2003} is $13.57$ for the $K$-band,
and the sky coverage is 414 square degrees.  Using this information,
we can estimate the maximum observational volume, $V_i$, for a galaxy
with an absolute magnitude of $M_i$.  Assuming that galaxy clustering
is negligible on the scale of this maximum volume, we can write the
number of galaxies with this absolute magnitude as
\begin{equation}
n_{a}(M_i)={\rm Integer}(\Phi_i 
\Delta M_i V_i)\,, 
\end{equation}
where $\Phi_i\equiv \Phi(M_i)$ is the luminosity function 
at an absolute magnitude $M_i$, and  
$\Delta M_i$ is the bin size. 
The logarithmic Poisson likelihood for a given model that predicts 
the luminosity function as $\Phi_i$ is 
\begin{equation}
\ln L=\displaystyle\sum\limits_{i=0}^k \left[n_a(M_i)\ln \left(\Phi_i \Delta M_i V_i\right) - \Phi_i \Delta M_i V_i
- \Gamma \left(1+ n_a(M_i)\right)\right],
\end{equation}
where the summation is over all the magnitude bins, and $\Gamma$ is 
the Gamma function. 

However, when there is faint-end incompleteness, one should not expect
$\Phi_i V_i$ galaxies in the faint-end bins. We define the completeness
fraction of a magnitude bin $i$ as the ratio between the number of the
observed galaxies and the total number of galaxies in a volume limited
sample, i.e. $p_i={\Phi_{{\rm obs},i}/\Phi_i}$, where $\Phi_{{\rm obs},i}\Delta M_i$ is
the observed number density in the $i$th bin, while $\Phi_{i}\Delta
M_i$ is the actual number density.  Thus, if a model predicts $N_i$
galaxies in the magnitude bin, then the number of galaxies to be
observed is $p_i N_i$.  \citet{Bell2003} estimated the incompleteness
in the $K$-band at the faint end using SDSS data, and found that for a
complete sample the slope of the $K$-band luminosity function at
$M_{\rm K} -5\log_{10} h>-21$ could be as steep as $-1.33$, compared
to the slope of $-0.93$ obtained directly from the data
\citep{Cole2001}.  This suggests that the completeness ratio may be
approximated by a power law in luminosity, or equivalently, an
exponential function of absolute magnitude. Following this
observational result, we assume that $p_i$ is unity for $M_{\rm K}
-5\log_{10} h<-21$ but decreases toward the faint end as
\begin{equation}
p_i=10^{-\alpha_{\rm IN}(M_{{\rm K},i} + 21 - 5 \log_{10} h)}\,,
\end{equation}
where $\alpha_{\rm IN}$ is a constant describing how fast the
incompleteness changes with magnitude, and its value is equal to the
difference in the faint-end slope between the incomplete and complete
samples. Because the exact value of $\alpha_{\rm IN}$ is uncertain, we
treat is as a free parameter with a prior distribution based on the
result of \citet{Bell2003}. The observations suggest that the
luminosity function at the faint end, if fitted by a power law, may
take any slope between $-0.93$ and $-1.33$. Hence, we assume that the
angle in logarithmic space
between the power law of a complete luminosity function and the
directly observed power law of $L^{-0.93}$ has a uniform probability
distribution: $\arctan(\alpha_{\rm IN})$ is uniformly
distributed between 0 and $\arctan(-0.93)-\arctan(-1.33)=0.177$.
Thus, if the predicted faint-end slope is $-1.33$, then the faint-end
slope to be observed could be a random value anywhere between $-1.33$
and $-0.93$.

Taking into account the incompleteness, the logarithmic 
Poisson likelihood for the luminosity function is
\begin{equation}
\ln L= \displaystyle\sum\limits_{i=0}^k \left[n_{a}(M_i) \ln \left(p_i \Phi_i M_i V_i \right) -  p_i \Phi_i M_i V_i -  \Gamma(1+n_{a}(M_i))\right]. 
\end{equation}
Thus, the Poisson likelihood is not only a function of the model
parameter vector $\theta$ but is also a function of 
$\alpha_{\rm IN}$. We treat $\alpha_{\rm IN}$ as a 
parameter in the inference and marginalise over it to compute the
observables using equation (\ref{eq-ppd}).

\section{Constraints on the Model Parameters}
\label{sec:constraint}

To obtain samples from the posterior distribution, we run
the Tempered Differential Evolution MCMC algorithm with 256 chains in
parallel.  We choose $T=9$ for the initial run and
  obtain convergence in 4500 iterations.  The Markov chain
broadly explores the parameter space in every dimension and converges
to real modes. We then resume the simulation from the
state at the 4500th iteration with $T=1$ to sample the
true posterior.  This procedure
  accelerates the convergence at the fiducial level with
  $T=1$. To achieve good mixing, we set a high maximum temperature,
$T_{\rm max}=128$ for the first powered-up level and $T_{\rm
  max}=1024$ for the fiducial level, for the tempering steps, which
occur for every 21 regular Differential Evolution steps. We stop the
simulation after 8000 iterations. Our Gelman-Rubin test finds that the
chains converge after 3000 iterations and we identify 14 outlier
chains.  We include the chain states of the last 3500 iterations,
847,000 states in total, to summarise the posterior. The
auto-correlation length is about 20, implying that there are
approximately 40,000 independent chain states.

\subsection{The posterior distribution}

\begin{figure}
    \begin{center}
      \epsfig{file=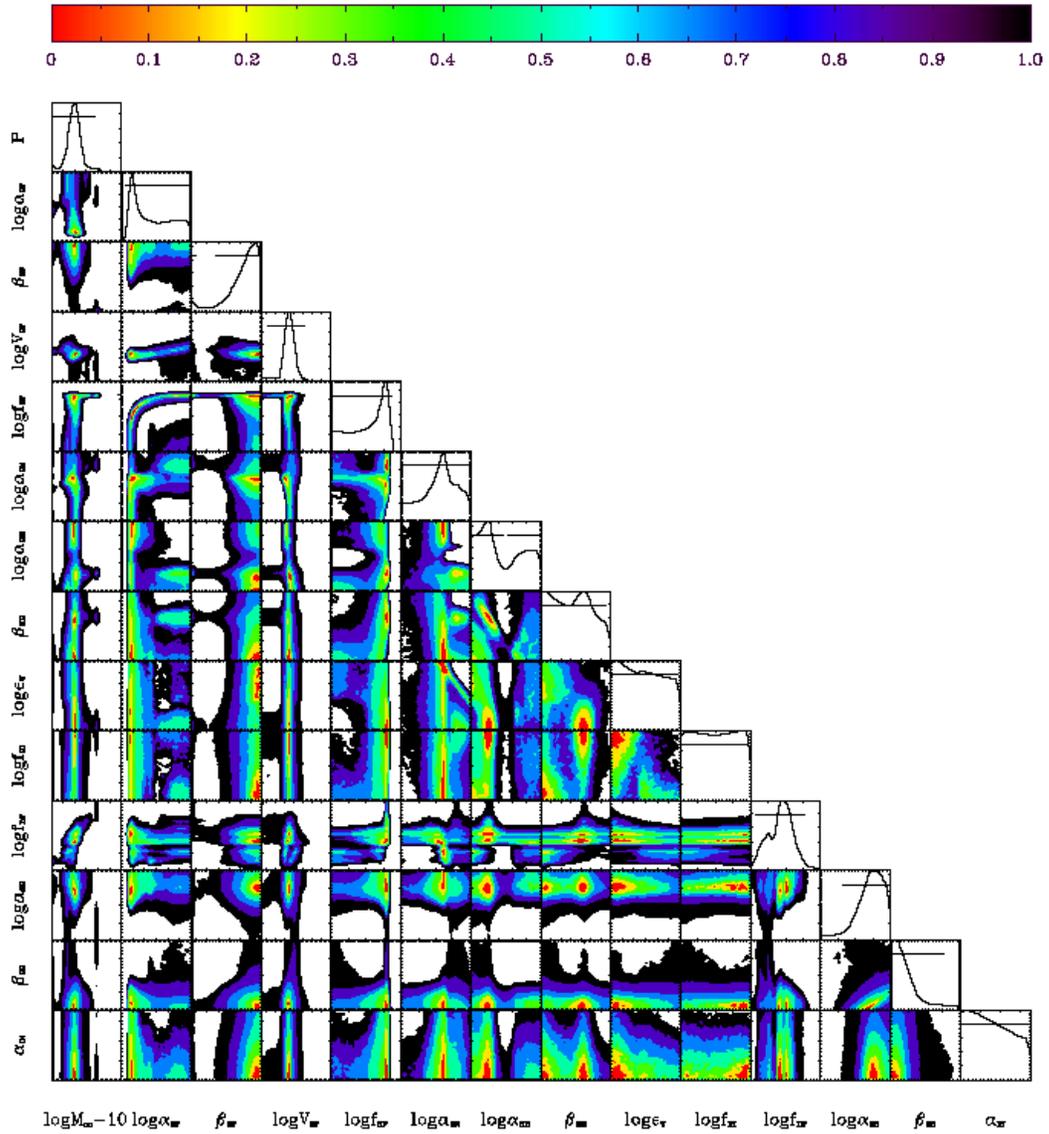, scale=0.8}
    \end{center}
\caption{\large
The 2-D and 1-D marginalised posterior probability density 
distributions for the 14 free parameters. 
}\label{fig:post_cb07}
\end{figure}

Figure \ref{fig:post_cb07} shows the marginalised posterior
distribution of the 14 parameters of the model family in question.
The posterior preserves many of the features that we saw in
\citet{Lu2011a}, which used a
synthetic version of the galaxy stellar mass function as the
observational constraint.  For example, the degeneracies between
$\Sigma_{\rm SF}$ and $\alpha_{\rm SF}$ and between $\alpha_{\rm RH}$
and $\beta_{\rm RH}$, and the strong constraint on the parameter
$V_{\rm SF}$, are clearly seen in both posterior distributions.  In
addition, the power indices $\beta_{\rm SF}$ and $\beta_{\rm RH}$ are
constrained to have large values of about 10, again similar to the
results obtained in \citet{Lu2011a}.  These similarities are not
surprising, since the stellar mass function used in \citet{Lu2011a} is
derived from the K-band luminosity function of galaxies. However, the
posterior distribution for some of the parameters obtained here
differs significantly from that obtained in \citet{Lu2011a}. For
example, in the stellar mass function constrained posterior, the
cooling cutoff mass, $M_{\rm CC}$, and the coefficient for the
dynamical friction timescale, $f_{\rm DF}$ are strongly degenerate and
bimodal: a model could have a large cooling cutoff mass, which
implies weak AGN feedback, if the dynamical friction timescale was
long. However, using the K-band luminosity function constraint,
models with very large $M_{\rm CC}$ are
  strongly disfavoured. 
The marginalised posterior only shows one dominant mode
with lower $M_{\rm CC}$ and $f_{\rm DF}$ values. Similar changes also
occur in some other parameters: e.g. $\beta_{\rm SF}$ and $\beta_{\rm
  RH}$.  These differences largely owe to the different error models.
As described in \S\ref{sec:data}, the error model used in this paper
is realistic, and so the resulting posterior distribution is reliable,
unlike those used in \citet{Lu2011a}.

\begin{figure}
  \hfill
  \begin{minipage}[t]{.45\textwidth}
    \begin{center}
      \epsfig{file=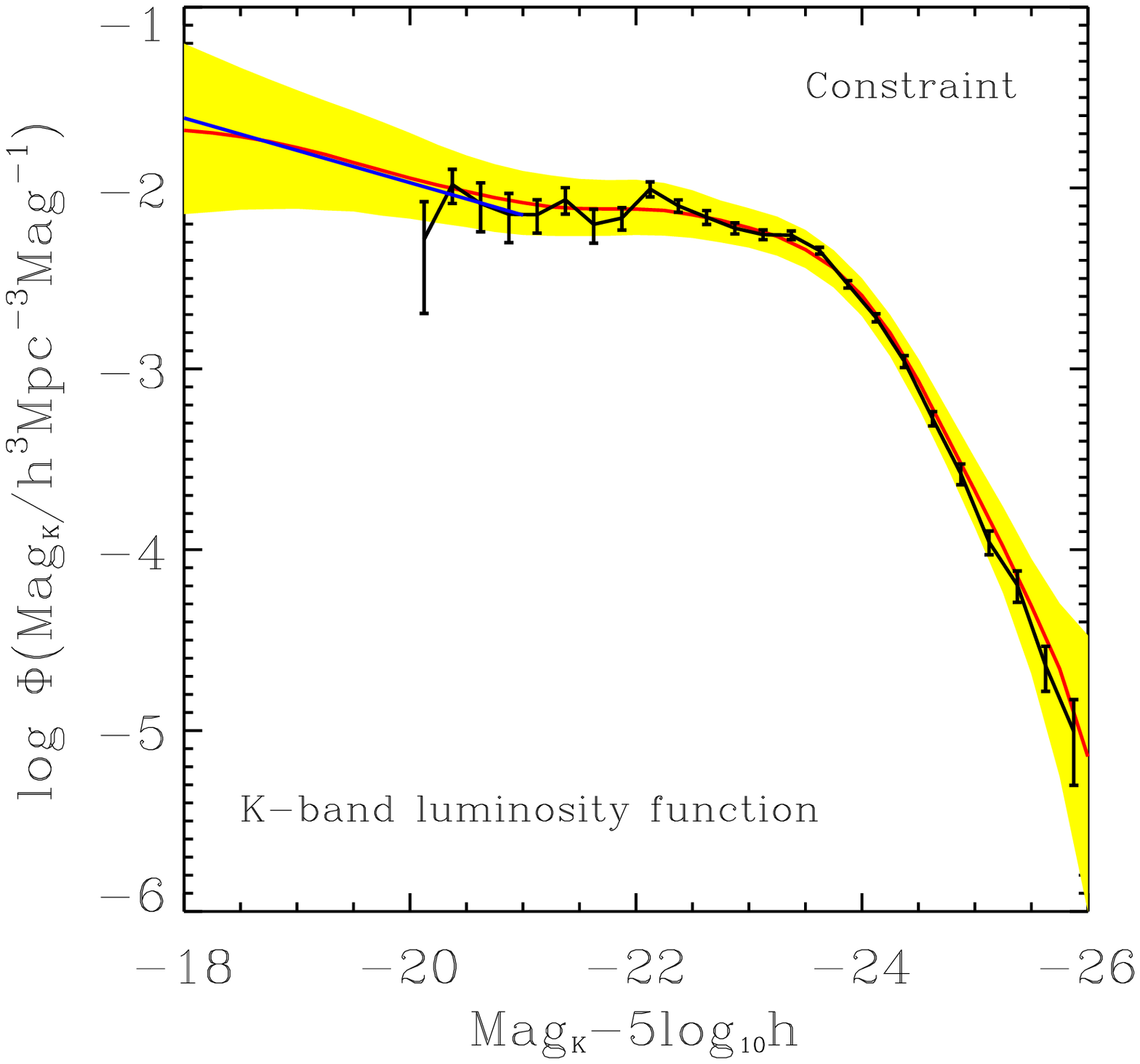, scale=0.4}
    \end{center}
  \end{minipage}
  \hfill
  \begin{minipage}[t]{.45\textwidth}
    \begin{center}
      \epsfig{file=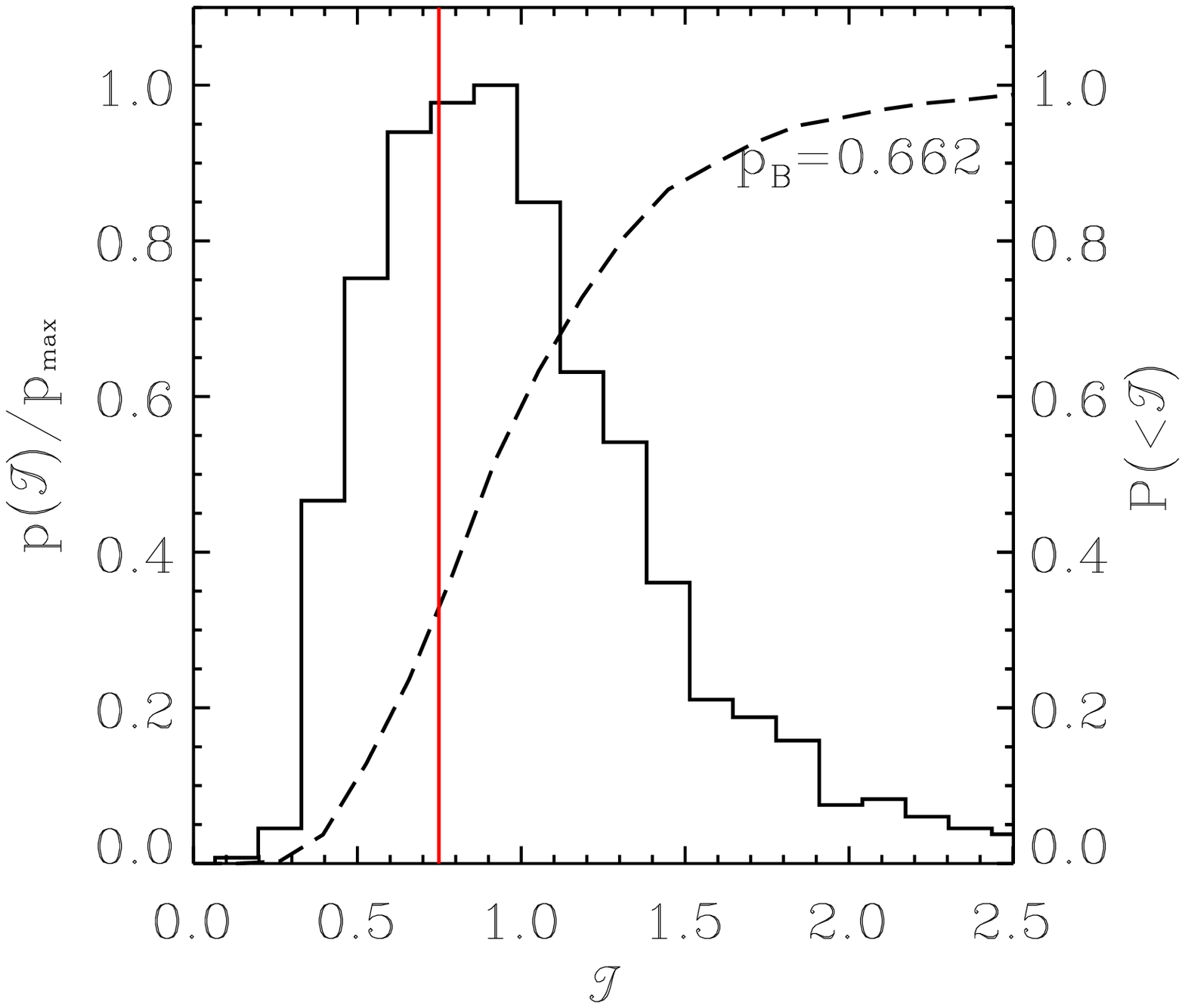, scale=0.4}
    \end{center}
  \end{minipage}
 \hfill
\caption{\large The left panel shows the Bayesian posterior predictions of the K-band
luminosity function at the present time. 
The posterior is constrained by the K-band luminosity function. 
The black solid line with error bars shows the observational data.
The yellow band encompasses the 95\% confidence range of the predictions
and the red line denotes the median value of the predictions.
The blue solid line shows the estimated faint end corrected for incompleteness \citep{Bell2003}.
The right panel shows the posterior predictive distribution 
of the test quantity $\calT$  for the K-band luminosity function.
The red line marks the position of the observed luminosity function 
in the distribution.
}\label{fig:klf}
\end{figure}

Figure \ref{fig:klf} shows the predicted $K$-band luminosity function at $z=0$.
The solid black lines with error bars shown in the 
left panel are the observational results.
The blue line sketches the estimated faint end of the luminosity 
function  corrected for incompleteness \citep{Bell2003}.
The yellow bands encompass the 95\% confidence range 
of the predictions, while the red solid line is the median.  
Clearly, the model family considered here can accommodate 
the observed $K$-band luminosity function remarkably well.
This is also demonstrated clearly with the posterior predictive 
check (PPC) described in \S\ref{sec:bayes_check}, as is 
shown in the right panel of Figure \ref{fig:klf}
with the corresponding value, $p_B=0.662$, given in the panel.
Note that we only use all the magnitude bins with $M_{\rm K} -5\log_{10} h>-21$ 
to perform the PPC, as we have assumed that the faint end of the observed 
luminosity function is incomplete.   

\subsection{Comparison with other semi-analytic models}

Here, we compare our posterior distribution with the model parameters used in
other SAMs.  As detailed in \citet{Lu2011a}, our models of star
formation and feedback encompass many published models as
subsets. For example, our star formation model works in the same way
as the Galform model \citep{Cole2000, Bower2006} except that we
include a variable cold gas surface density threshold for star
formation. Our star formation model can also be reduced to the Munich
model \citep{Croton2006} by setting the parameter $\beta_{\rm SF}$ to 0.  
Our SN feedback model is similar to the Munich model but allows more
parameters to vary. Inspecting our posterior distribution, one can
find that some of the modes that we identify are broadly consistent
with those found by other studies. \citet{Henriques2009} found that
$\epsilon_{\rm disc}$ in the model proposed by \citet{Croton2006},
which corresponds to $\alpha_{\rm RH}$ in our model, is required to be
as high as about 10. The SN feedback parameters in
\citet{Somerville2008}, $\epsilon_{\rm SN}$ and $\alpha_{\rm RH}$,
which correspond to our $\alpha_{\rm RH}$ and $\beta_{\rm RH}$, were
tuned to be 2 and 1.3. These values are right on the ridge of the
marginalised posterior distribution of those dimensions.
\citet{Bower2010} found that the normalisation for the star formation
efficiency is as low as about 0.003, which is also similar to
our mode for $\alpha_{\rm SF}$.  The
dynamical friction time scale coefficient obtained here is in broad
agreement with the merging time scales adopted in other
SAMs. The posterior distribution shows that our $f_{\rm DF}$ is 
a few times larger than the corresponding coefficients in the Durham 
model and the Munich model. However, those models use the satellite galaxy's
mass whereas we use its halo mass to compute the timescale, so
their model parameters actually agree well with the
posterior for the parameter in our model.  Our prescription for
merger-triggered starbursts is very similar to the Munich model
\citep{Croton2006} and the Somerville model \citep{Somerville2008},
and the model parameters they adopted are contained in the modes of
the posterior we obtain.  All these similarities between our model and
other existing models and the consistency between the posterior modes
we obtain and the parameter values adopted in other SAMs imply that,
although our inference is based on a specific model family, our
inference may hold for other SAMs.

\section{Model Predictions versus Observational Data}
\label{sec:prediction}

In the following we concentrate on a number of important
observables and examine how our model
predictions compare with available observations. To obtain the
predictions using equation (\ref{eq-ppd}), we randomly select 1000
samples from posterior distribution. For presentation, we
also randomly select 8 parameter sets from the posterior sample and
plot their individual predictions for some observables.

\subsection{The local HI mass function}

\begin{figure}
    \begin{center}
      \epsfig{file=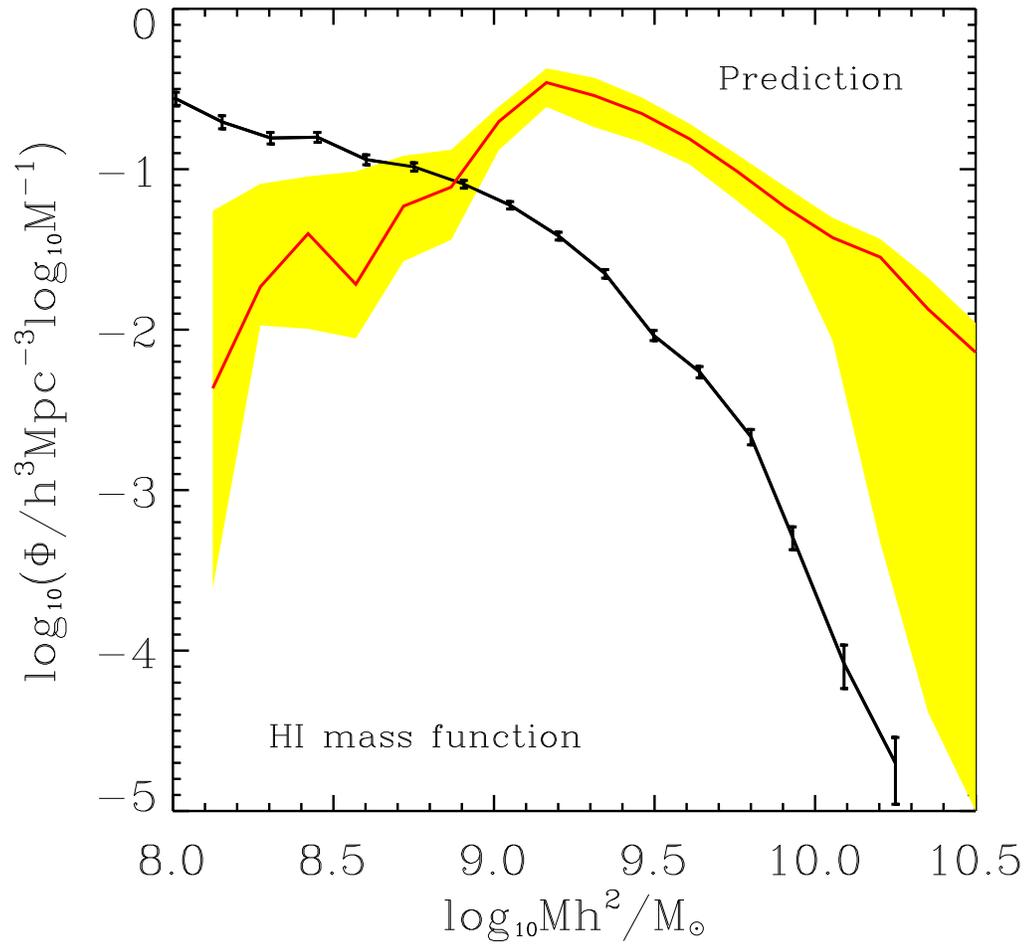, scale=0.8}
    \end{center}
\caption{\large
The posterior predictions of the HI gas mass
function at the present time compared with the HI gas mass function 
of local galaxies obtained by \citet{Zwaan2005a}. 
The black solid line with error bars denotes the observational data.
The yellow bands encompasses the 95\% confidence range of the predictions
and the red lines denote their median value. The downturn at low masses
is caused by resolution effects.
}\label{fig:h1mf}
\end{figure}

Figure \ref{fig:h1mf} shows the HI gas mass function of 
galaxies at $z=0$ predicted by our constrained model 
compared with the observational data of the HI gas mass 
function of local galaxies obtained by \citet{Zwaan2005a}. 
To convert the cold gas mass in the model to 
the mass of cold hydrogen, we multiply the predicted cold gas mass 
by a factor $\beta=0.74$, 
the mass fraction of hydrogen in neutral gas with 
the rest consisting of helium (He) and a minor fraction of heavier
elements \citep{Obreschkow2009a}. Furthermore, since 
part of the cold hydrogen gas may be in ${\rm H}_2$ instead of in 
HI, the contribution of ${\rm H}_2$ has to be considered
when comparing our model predictions with the 
observational data. Unfortunately, our current model does not 
trace the formation of ${\rm H}_2$ so we use a simple 
model to include the contribution of ${\rm H}_2$.
According to the observational results of \citet{Keres2003}, the
total mass density of ${\rm H}_2$ in the local universe is about 0.4 
times the that of the HI gas. Assuming that the ${\rm H}_2$ mass in a 
galaxy is proportional to its HI mass, we obtain the HI mass 
by multiplying the predicted cold hydrogen mass by 
$1/1.4$. The model predictions shown in Figure \ref{fig:h1mf} 
are, therefore, the cold gas mass function predicted by the model multiplied 
by $0.74/1.4$. A similar conversion factor is adopted by \citet{Power2010}.  
 
The predicted cold gas mass function
is higher than the observed function by a factor of more than 5. The turn down
at low HI masses is artificial and results from the mass resolution limit of
the halo merger trees used here, which is $4.5\times10^9 \msun$.
Not surprisingly, our PPC indicates a significant difference
between the model predictions and the data, with $p_B=0.000$.  This
occurs because the total fraction of baryons in stars is small
compared to the total amount of gas that can cool, and the feedback is
not sufficient to remove the cold gas from the galaxies.
Consequently, a large amount of the cooled gas has to remain as cold
gas in galaxies. As discussed in \S\ref{sec:constraint}, the posterior
distribution of the parameters characterising the efficiency of
supernova feedback is already pushed to the extreme, suggesting that
the current model family may not be able to accommodate the observed
stellar and cold gas contents of local galaxies simultaneously. 
The problem of overpredicting the cold gas mass function is
not only in the model family we consider in the present paper.
Recently, \citet{Wang2011} also showed the same problem in
various models with different parameterisations of star formation. 
As first pointed out by \citet{Mo2005}, this is a generic problem for
current models of galaxy formation that use supernova feedback to
reheat and eject gas from galaxies.  Either supernova feedback is
severely underestimated, or some other process might be responsible for
preventing gas from being accreted by galaxy halos in the first place,
such as preheating \citep{Mo2002, Mo2005, Lu2007}. In a forthcoming
paper, we will use both the galaxy luminosity function and the HI mass
function as joint observational constraints to study their
implications for star formation and feedback.

\subsection {The Tully-Fisher relation}

\begin{figure}
  \begin{center}
    \epsfig{file=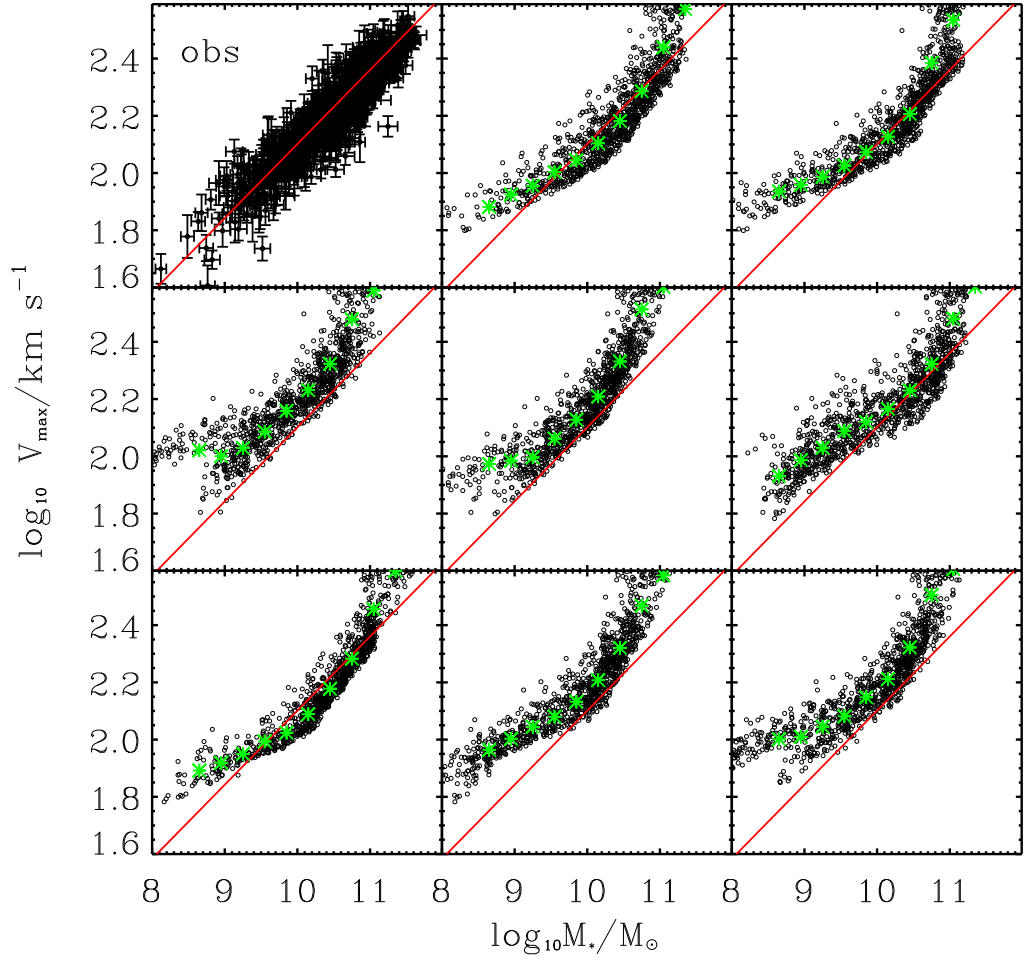, scale=0.8}
  \end{center}
\caption{\large The stellar mass Tully-Fisher relation predicted by 8 models randomly selected 
from the posterior compared with data from \citet{Dutton2010} shown in the upper-left panel. 
The red line denotes a fit to the observational data given by \citet{Dutton2010}.
}\label{fig:tf}
\end{figure}

 Figure \ref{fig:tf} shows the Tully-Fisher relations predicted by 
8 randomly selected models from the posterior distribution compared with 
the observationally derived data from \citet{Dutton2010}.
Following \citet{Dutton2010}, we plot the maximum rotation 
velocities ($V_{\rm max}$) of galaxies versus their total stellar 
masses. As shown in  \citet{Dutton2010}, the observed Tully-Fisher 
relation follows a simple power law, 
\begin{equation}
\log \left( { V_{\rm max} \over {\rm km s^{-1}}}\right) 
= 2.179 + 0.259 \log \left( {M_* \over 10^{10.3} \msun}\right)\,,
\end{equation}
and the 1$\sigma$ uncertainty in the zero point is about 0.005.  For
comparison, this power law is shown as the straight line in each
panel.  To mimic the observations, we only select disk-dominated central
galaxies at $z=0$ that have bulge/total stellar masses smaller than
1/10. Unlike the stellar mass, the
maximum rotation velocity is not a direct prediction of our model.  As
a simple model, we use the peak circular velocity of the host dark
matter halo as a proxy for $V_{\rm max}$, ignoring any effects owing to
the baryonic component.  The halo peak circular velocity is computed
according to the circular velocity---halo mass relation obtained by
\citet{Klypin2010} from $N$-body simulations:
\begin{equation}
V_{\rm max}=2.8\times10^{-2} \left({\mvir \over \msunh}\right)^{0.316} {\rm km/s}\,,
\end{equation}
where $V_{\rm max}$ is the halo peak circular velocity and $\mvir$ is
the halo virial mass. With all these assumptions, the predicted
Tully-Fisher relation is independent of the
bulge/total ratio adopted to select disk-dominated galaxies.

As demonstrated in previous investigations, the amplitude of the
Tully-Fisher relation can be reproduced roughly in the current CDM
model if halo contraction owing to the growth of a central galaxy is
ignored \citep[e.g.][]{Choi2006, Dutton2007}. Our results shown in
Figure \ref{fig:tf} are consistent with these investigations.
However, unlike the observed power law relation, the predicted relations
are concave upward. The figure
illustrates that differences between the model predictions and
the data are significant. The green asterisks show the averaged $\log
V_{\rm max}$ in ten $\log M$ bins. We use the binned results to
perform a $p$-value PPC and find $p_B=0.005$, which suggests a poor
fit to the data.

This predicted curved shape is a direct consequence of the halo
mass-stellar mass relation found for central galaxies, which shows
that the halo to stellar mass ratio is the lowest for halos with
masses $\sim 10^{12}\msun$ and goes up towards both the low- and
high-mass ends \citep[][]{Yang2003, Yang2008}. This curved shape is
also seen in other semi-analytic models
\citep[e.g][]{Croton2006,Benson2010d}.  There are at least two possible 
explanations for this mismatch.  First, the observational relation may be
subject to selection effects that are not included in the model
predictions. For example, many of the galaxies at the low- and
high-mass ends might not be spirals observationally. 
Second, including halo-galaxy interactions
may reduce variations in the predicted Tully-Fisher relation. Indeed, as shown
in \citet[][]{Mo2000}, the interaction between the dark matter halo and
the disk as given by adiabatic contraction can reduce the scatter in
the Tully-Fisher relation produced by a variation in the baryon
fraction in galaxies, making the predicted Tully-Fisher relation
closer to a power law.  Unfortunately, this effect will also boost
$V_{\rm max}$, changing the overall amplitude of the predicted
relation.  Our results show that the boost has to be weak for the model
predictions to match the overall Tully-Fisher amplitude. It is unclear
if a consistent model can be found along these lines, or if galaxy
halos have density profiles shallower than those predicted by CDM models
\citep[e.g.][]{Mo2000, Weinberg2002}, or if interactions with the
baryonic component can make a halo profile shallower \citep[e.g.][]{Binney2001, El-Zant2001,Mo2004}.

\subsection {The colour distribution}

\begin{figure}
   \begin{center}
     \epsfig{file=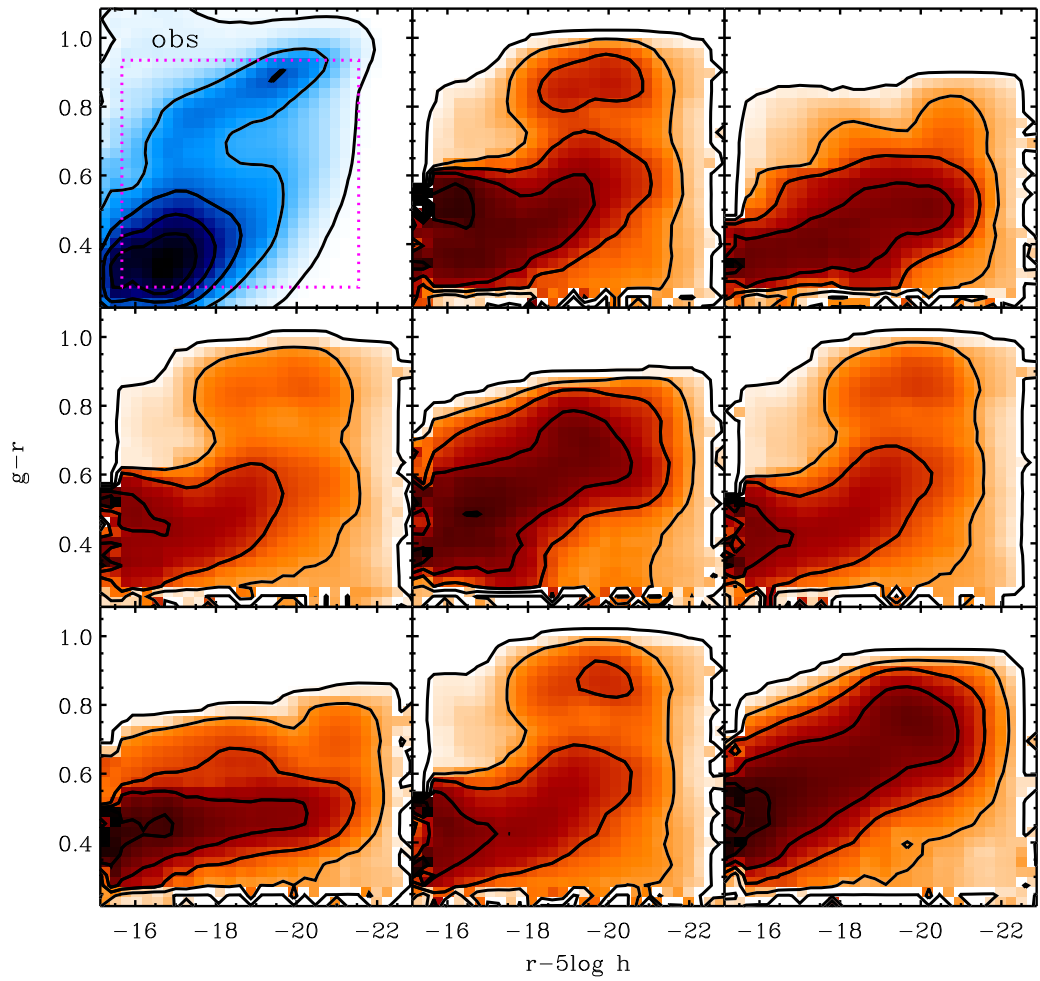, scale=0.8}
   \end{center}
\caption{\large The colour - magnitude diagram predicted by 8 models
randomly selected from the posterior compared with observational data
from SDSS (the upper-left panel).  The magenta dotted line encloses the
square region that is used to conduct a PPC.
}\label{fig:csmd}
\end{figure}

Figure \ref{fig:csmd} shows the distribution of galaxies in  the $g-r$
colour versus $r$-band magnitude plane derived from SDSS DR7 and the
same distribution predicted by 8 individual models randomly selected
from the posterior distribution.  Clearly, the model predictions are
diverse. Some models, such as the one shown in the middle of the upper
row, can reproduce the bimodal distribution seen in the $z\sim 0$
galaxy population. However, many other models do not predict any
bimodality, and the predicted colours may be bluer or redder than the observed
distribution.  We select the region of the diagram enclosed by the
magenta square, and divide the region into $25 \times 25$ bins. We
renormalise the colour distribution in each magnitude bin for all the
prediction samples and the observational data, and use these bins to
perform a $p$-value PPC with a result that $p_B=0.000$.  These results
suggest that the model parameters constrained by the $K$-band
luminosity function alone do not provide significant constraints on
the colour---magnitude relation.  This also implies that the observed
colour distribution can provide a constraint that is complementary to
the one provided by the luminosity (stellar-mass) function.

\subsection{The conditional stellar mass function}

\begin{figure}
    \begin{center}
      \epsfig{file=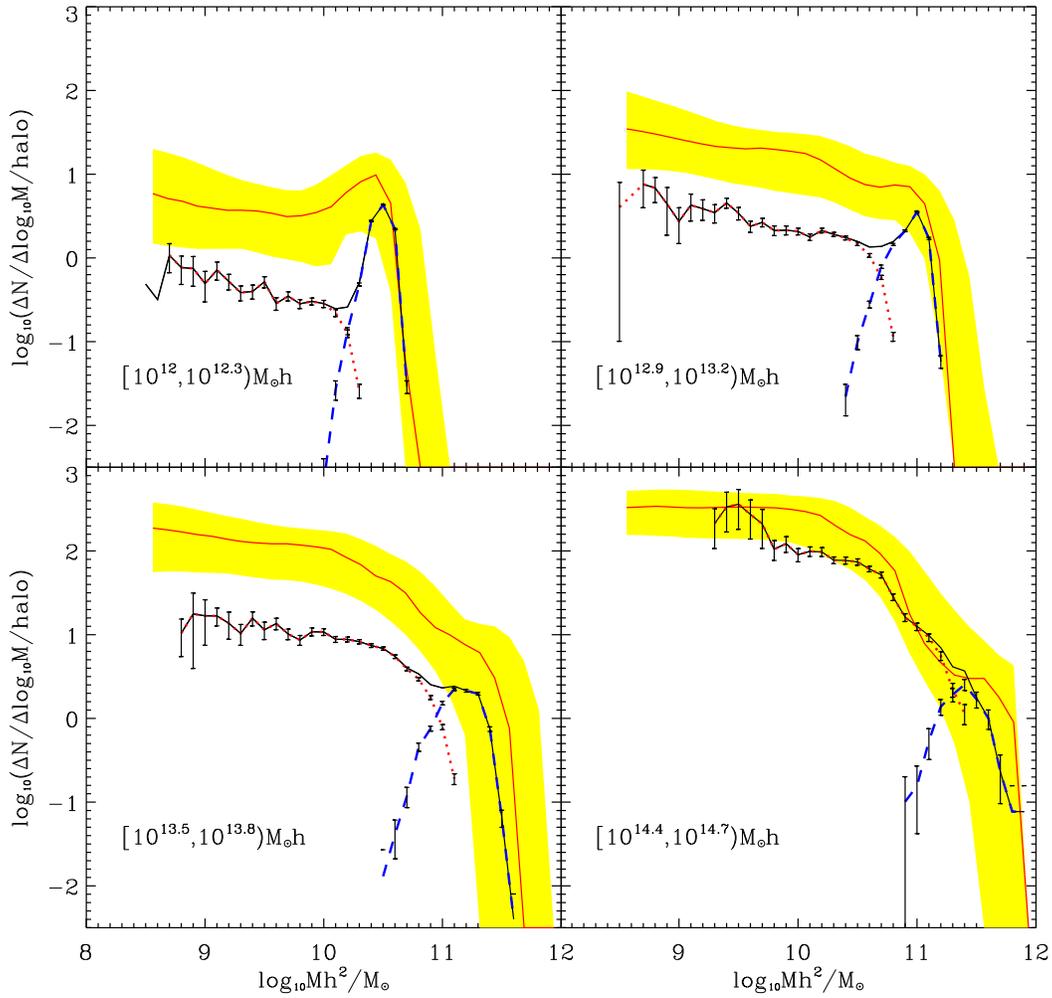, scale=0.8}
    \end{center}
\caption{\large
The Bayesian posterior predictions of the conditional galaxy stellar mass
functions for four halo masses at the present time.
The halo mass ranges are noted in each panel. 
The black solid lines with error bars denote
the observed CSMFs for all galaxies that 
reside in halos with the corresponding virial masses.
The blue dashed lines show the
CSMFs for central galaxies only and the red dotted lines show
that of satellite galaxies.
The yellow bands encompass the 95\% confidence range of the predictions
for the satellite galaxies and the red lines denote their median value.
}\label{fig:csmf}
\end{figure}

We also make predictions for the conditional stellar 
mass function (CSMF) of galaxies at $z\sim 0$, $\Phi(M_*|M_{\rm h})$, 
which is defined as the average number
of galaxies as a function of galaxy stellar mass
in host dark matter halos of a given mass. 
We compare our model predictions with the observed 
CSMFs given by \citet{Yang2008,Yang2009a}.
Our goal is to check whether the model family, which can
accommodate the observed total $K$-band luminosity function,  
is also able to accommodate the observed stellar contents in halos of 
different masses.
Following the presentations in Yang et al.,  
we obtain the CSMFs for the following four halo mass 
ranges: $10^{12}-10^{12.3}\msunh$, 
$10^{12.9}-10^{13.2}\msunh$, $10^{13.5}-10^{13.8}\msunh$, 
and $10^{14.4}-10^{14.7}\msunh$. 
The corresponding galaxy populations for those halo masses 
are modelled with
500, 300, 100 and 50 sampled halos, respectively. 
For each mass range, the halo samples are drawn from the halo 
mass function given by \citet{Sheth1999} and \citet{Sheth2001}. 
The results are shown in Figure \ref{fig:csmf}.
For the observational data, each CSMF is separated into 
two parts, the contribution of central
galaxies (defined to be the most massive galaxy 
in a group) and the contribution of satellite 
galaxies (all other galaxies in a group except the central). 
As one can see, halos of lower masses on 
average contain a smaller number of satellites, 
and so the central term is more prominent in the CSMF. 
For the model prediction we only present the total 
CSMF for each case, but it is worth noting
that our predicted CSMFs for the central galaxies match 
the observational results quite well. As one can see, however, 
the model significantly over-predicts the number of
satellite galaxies in low-mass halos. 
It is worth noting that the discrepancy in satellite galaxies 
does not contradict with the excellent fit to the K-band luminosity function, 
because the field luminosity function is dominated by central galaxies 
at all magnitudes, while the CSMF is more sensitive to the satellite 
galaxy population. 
Applying the  PPC described in \S\ref{sec:bayes_check} to the 
CSMFs, we obtain $p_B=0.002$, $0.002$, $0.002$ 
and $0.024$ for the four mass bins (from low-mass 
to high-mass), 
respectively, suggesting that the over-prediction
is significant for the three low-mass cases and 
marginally significant for the most massive case.    
This result, obtained by exploring a large parameter space, 
reinforces the finding of \citet{Liu2010} that the current SAMs 
cannot match the observed CSMFs in some halo-mass ranges, 
even though they are able to reproduce the total stellar 
mass function. This discrepancy suggests that some important 
physics governing the evolution of satellite galaxies,
such as tidal stripping and/or tidal disruption, 
should be included in the model \citep[e.g.][]{Yang2009, Liu2010}.
\citet{Kang2008} have shown that tidal stripping can effectively reduce 
the fraction of red satellite galaxies in their model to achieve a better agreement 
with the data. \citet{Kim2009} have demonstrated that including tidal disruption 
and satellite-satellite mergers in their model can improve the match to
galaxy clustering on small scales. 
In a forthcoming paper, we will use the 
observed CSMFs as constraints to infer implications 
for the evolution of satellite galaxies.  

\subsection{Redshift evolution of cold baryonic masses}

\begin{figure}
  \hfill
  \begin{minipage}[t]{.45\textwidth}
    \begin{center}
      \epsfig{file=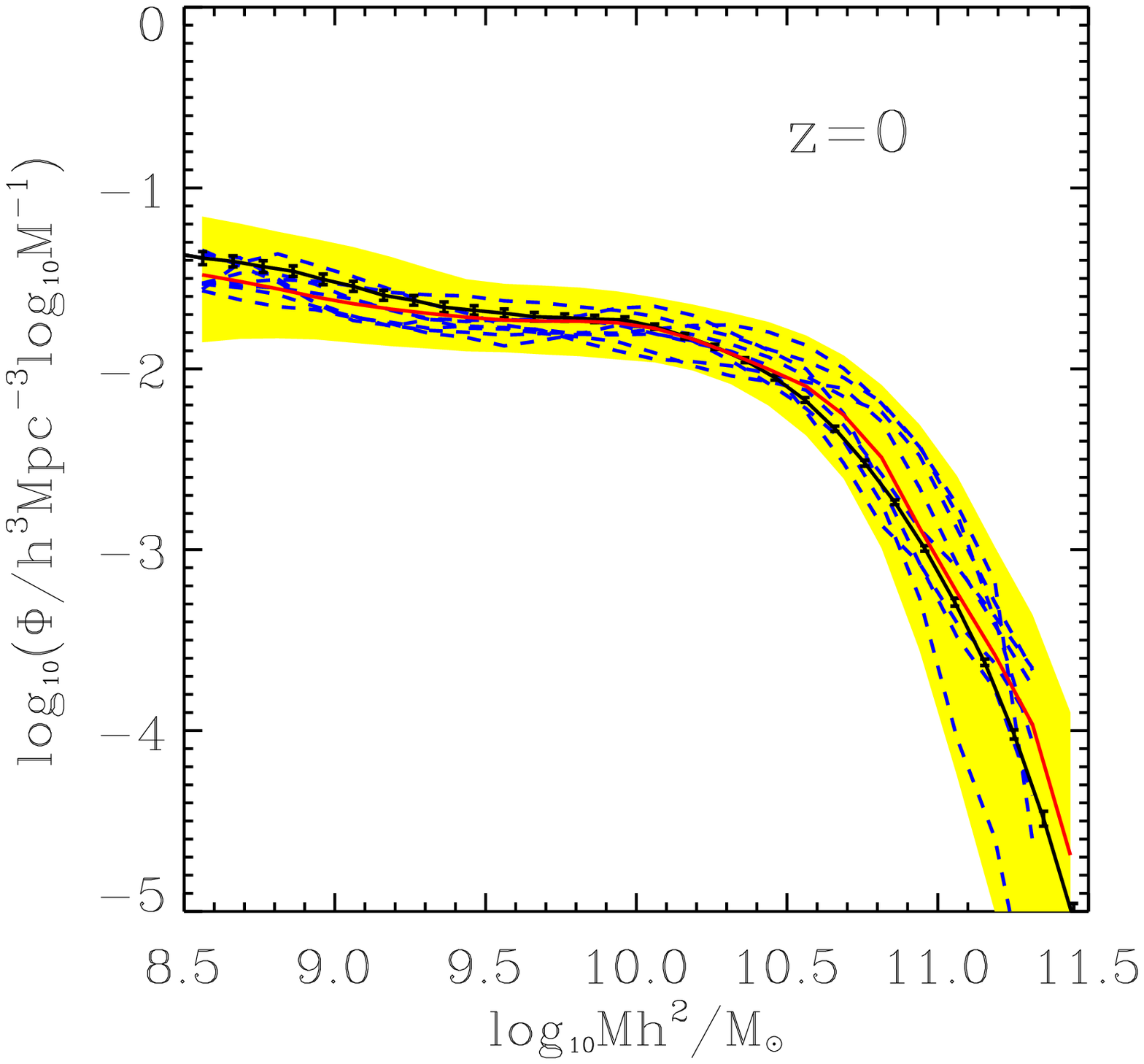, scale=0.4}
    \end{center}
  \end{minipage}
  \hfill
  \begin{minipage}[t]{.45\textwidth}
    \begin{center}
      \epsfig{file=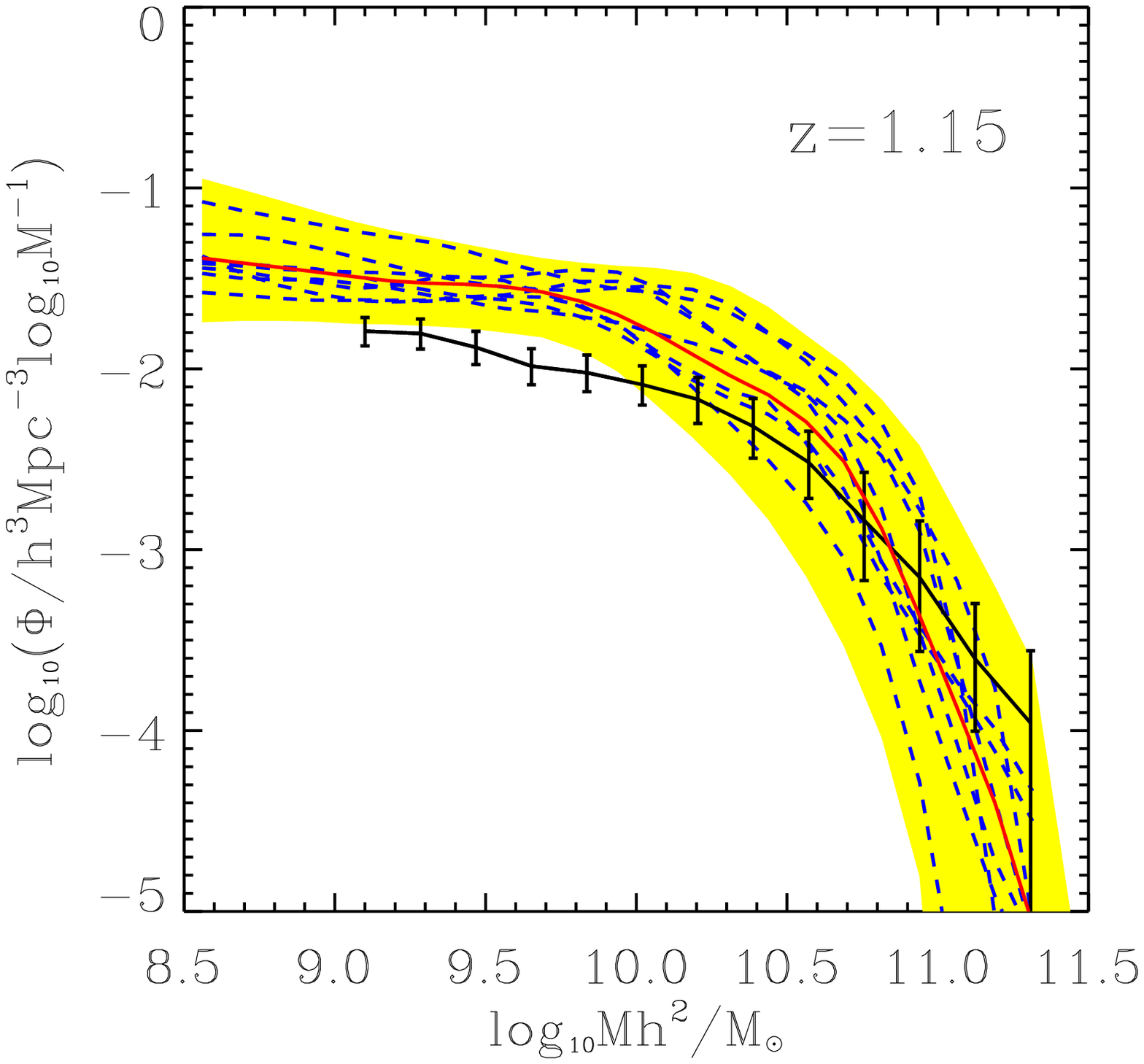, scale=0.4}
    \end{center}
  \end{minipage}
  \hfill

  \hfill
  \begin{minipage}[t]{.45\textwidth}
    \begin{center}
      \epsfig{file=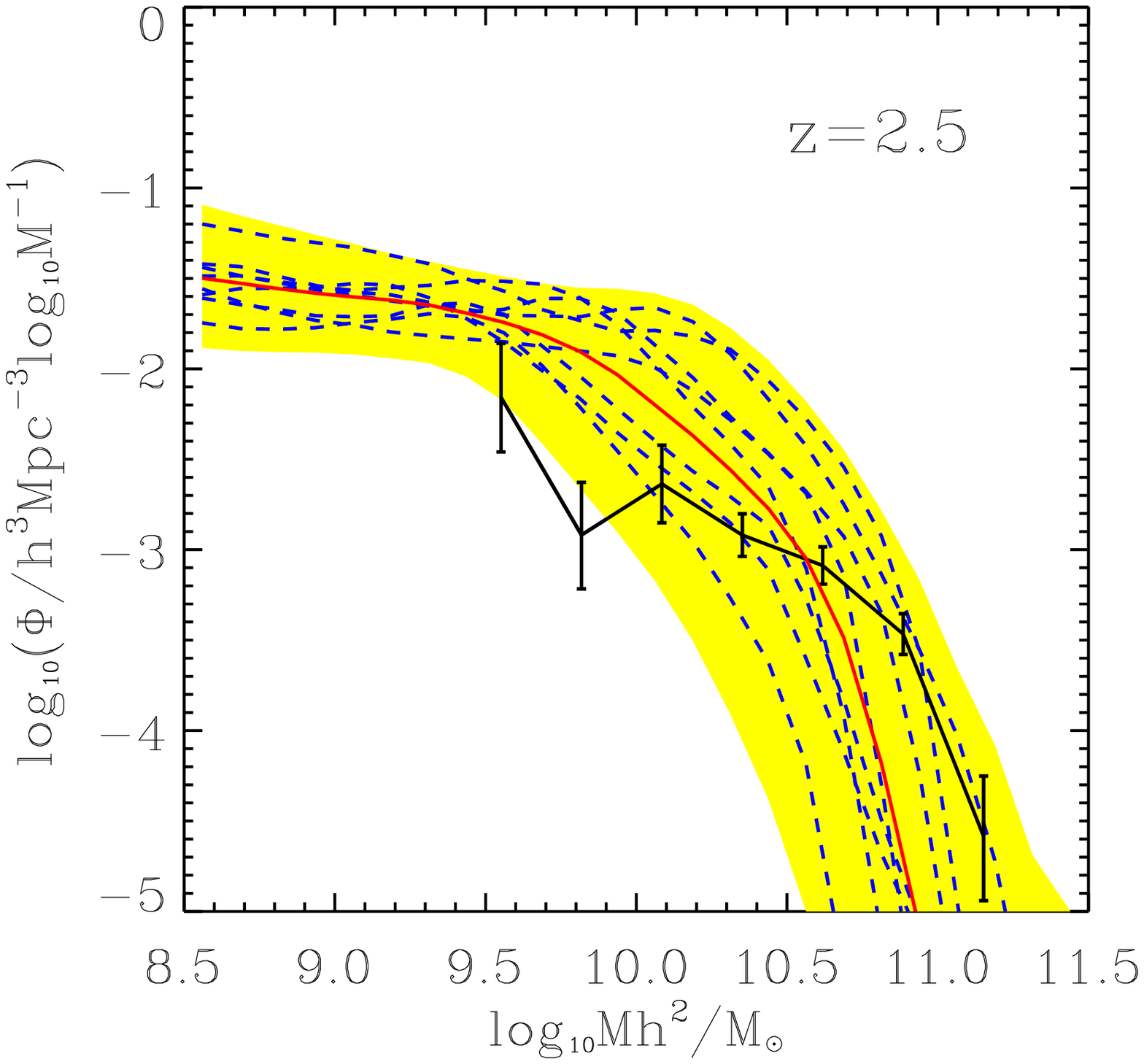, scale=0.4}
    \end{center}
  \end{minipage}
  \hfill
  \begin{minipage}[t]{.45\textwidth}
    \begin{center}
      \epsfig{file=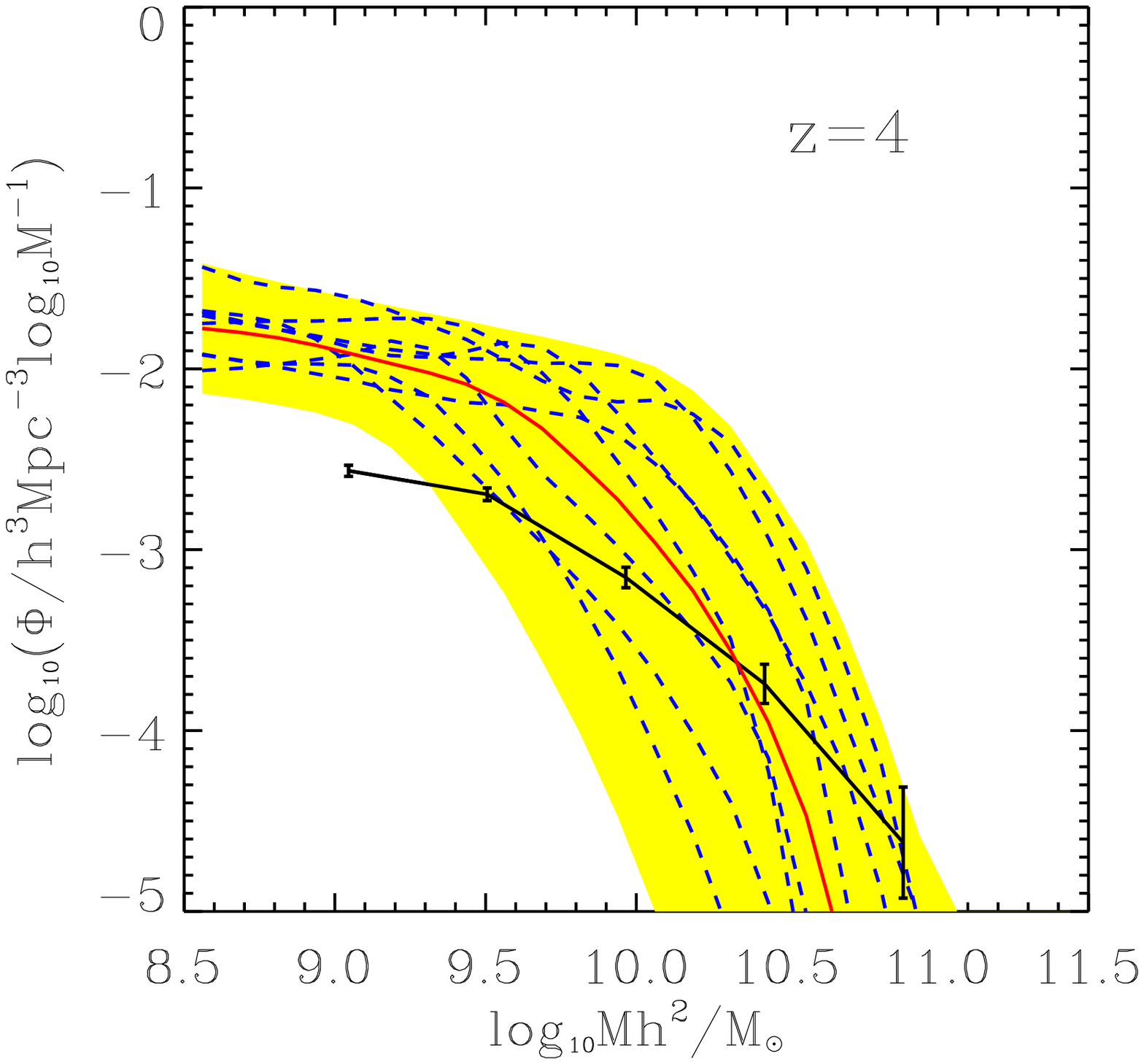, scale=0.4}
    \end{center}
  \end{minipage}
  \hfill

\caption{\large
The posterior predicted stellar mass functions at $z=$0, 1.15, 2.5 and 4.
The yellow bands enclose 95\% confidence range and the red line plots the 
median.  The blue dashed lines denote the predictions of 8 models randomly 
selected from the posterior sample.  
The data are the black solid lines with error bars. 
The stellar mass function for $z=0$ is from \citet{Li2009},
$z=1.15$ is from \citet{Perez-Gonzalez2008},
$z=2.5$ is from \citet{Marchesini2009}, and
$z=4$ is from \citet{Stark2009}.
}\label{fig:smfz}
\end{figure}

\begin{figure}
 \hfill
 \begin{minipage}[t]{.45\textwidth}
   \begin{center}
     \epsfig{file=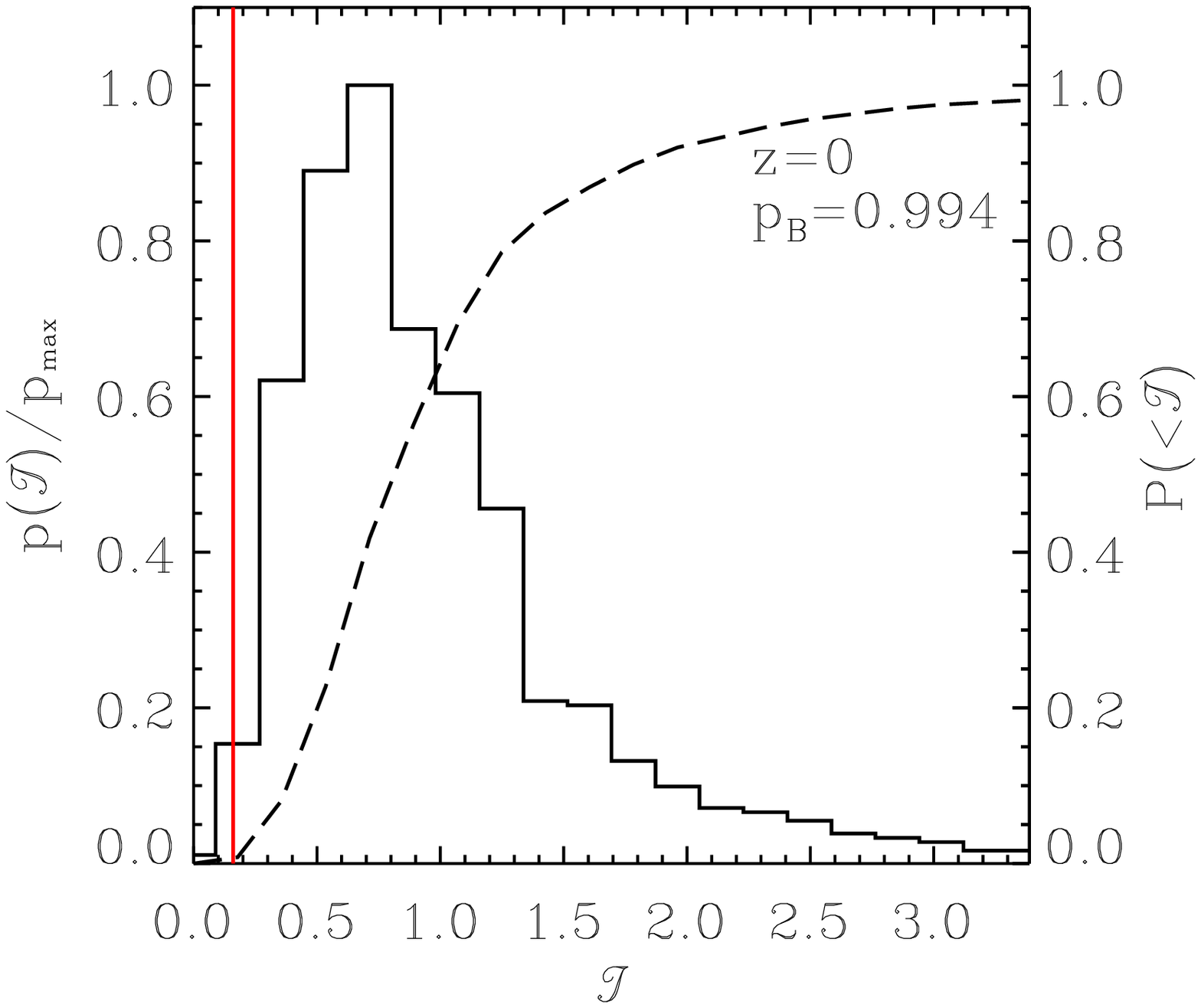, scale=0.4}
   \end{center}
 \end{minipage}
 \hfill
 \begin{minipage}[t]{.45\textwidth}
   \begin{center}
     \epsfig{file=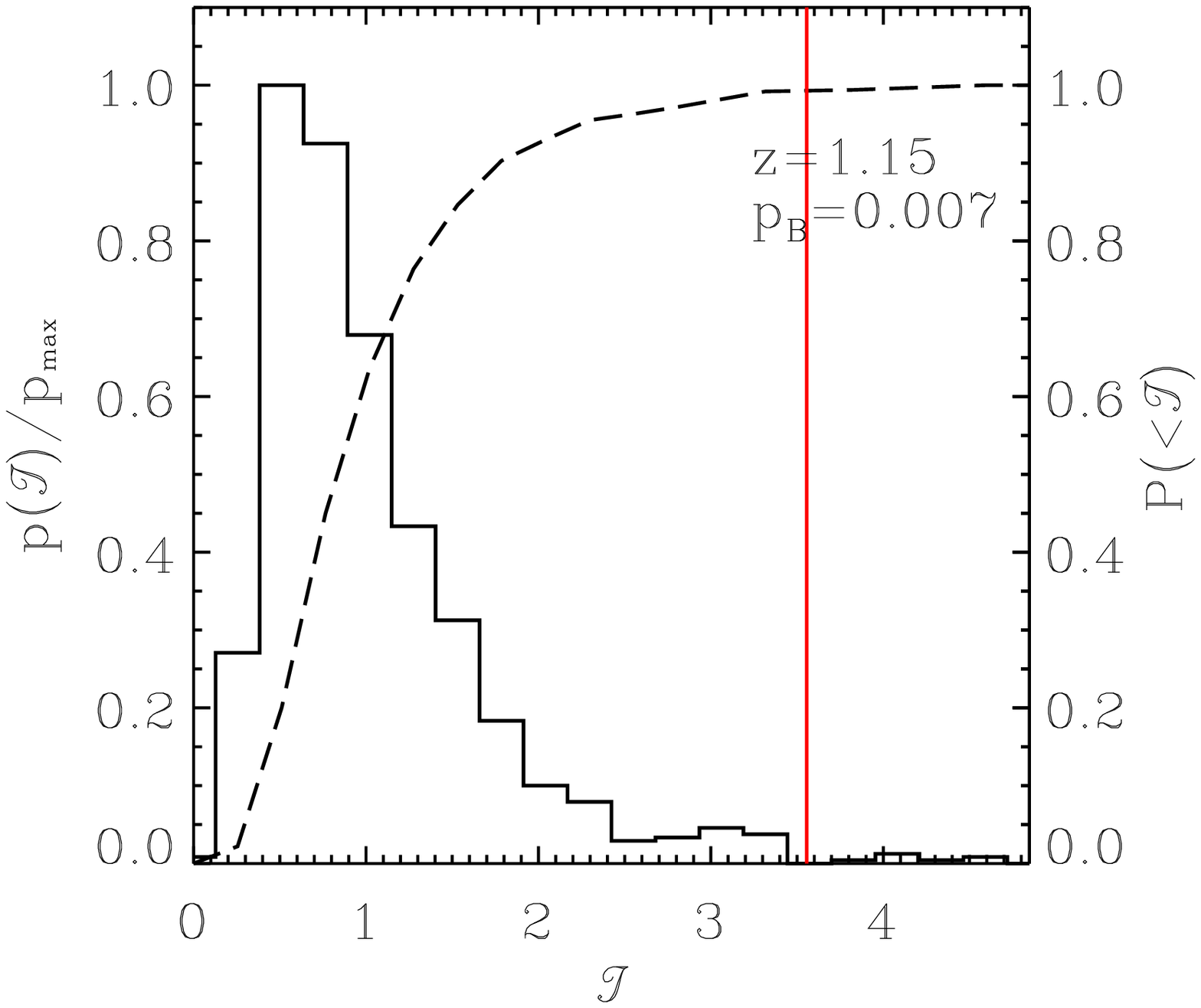, scale=0.4}
   \end{center}
 \end{minipage}
 \hfill

 \hfill
 \begin{minipage}[t]{.45\textwidth}
   \begin{center}
     \epsfig{file=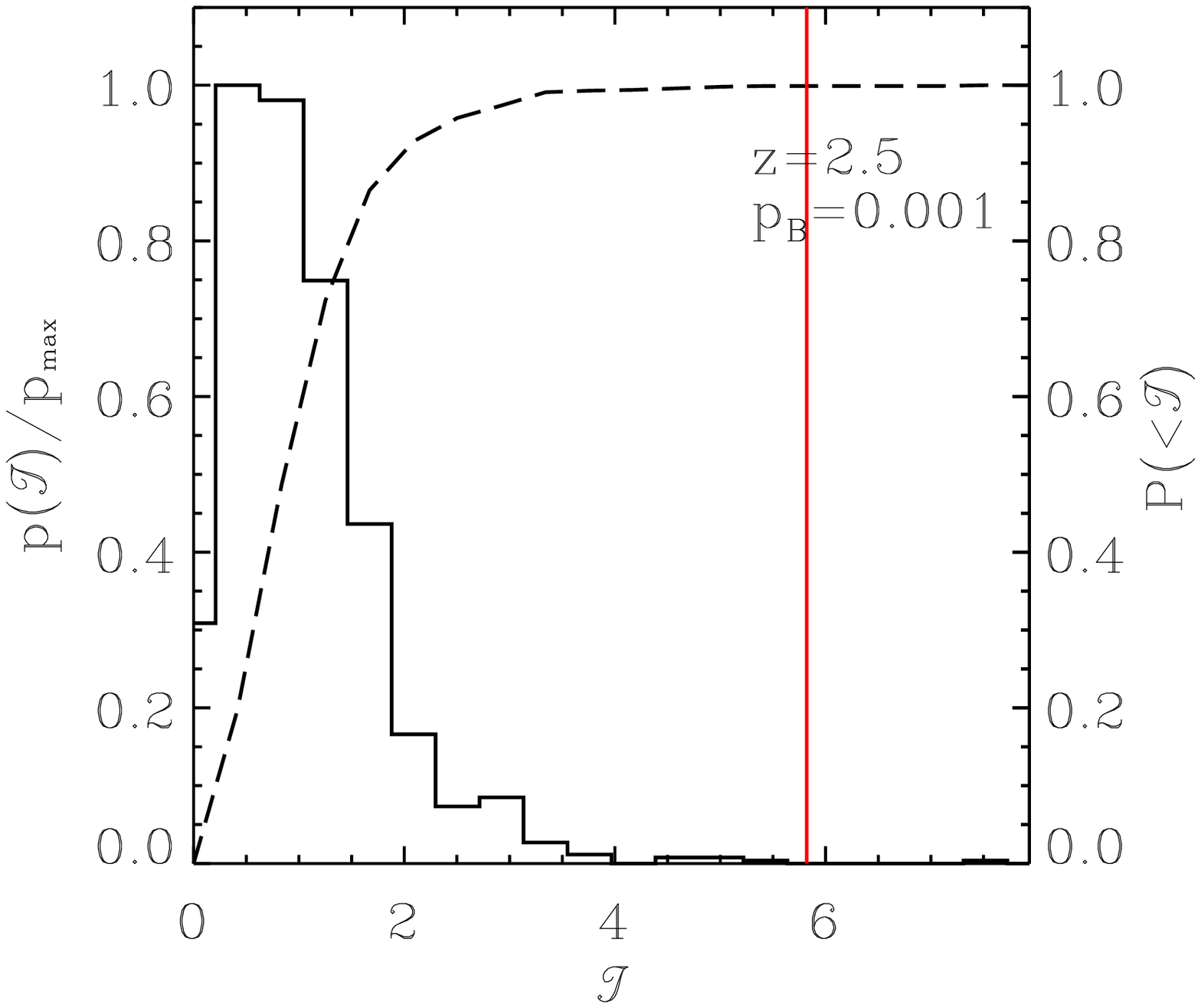, scale=0.4}
   \end{center}
 \end{minipage}
 \hfill
 \begin{minipage}[t]{.45\textwidth}
   \begin{center}
     \epsfig{file=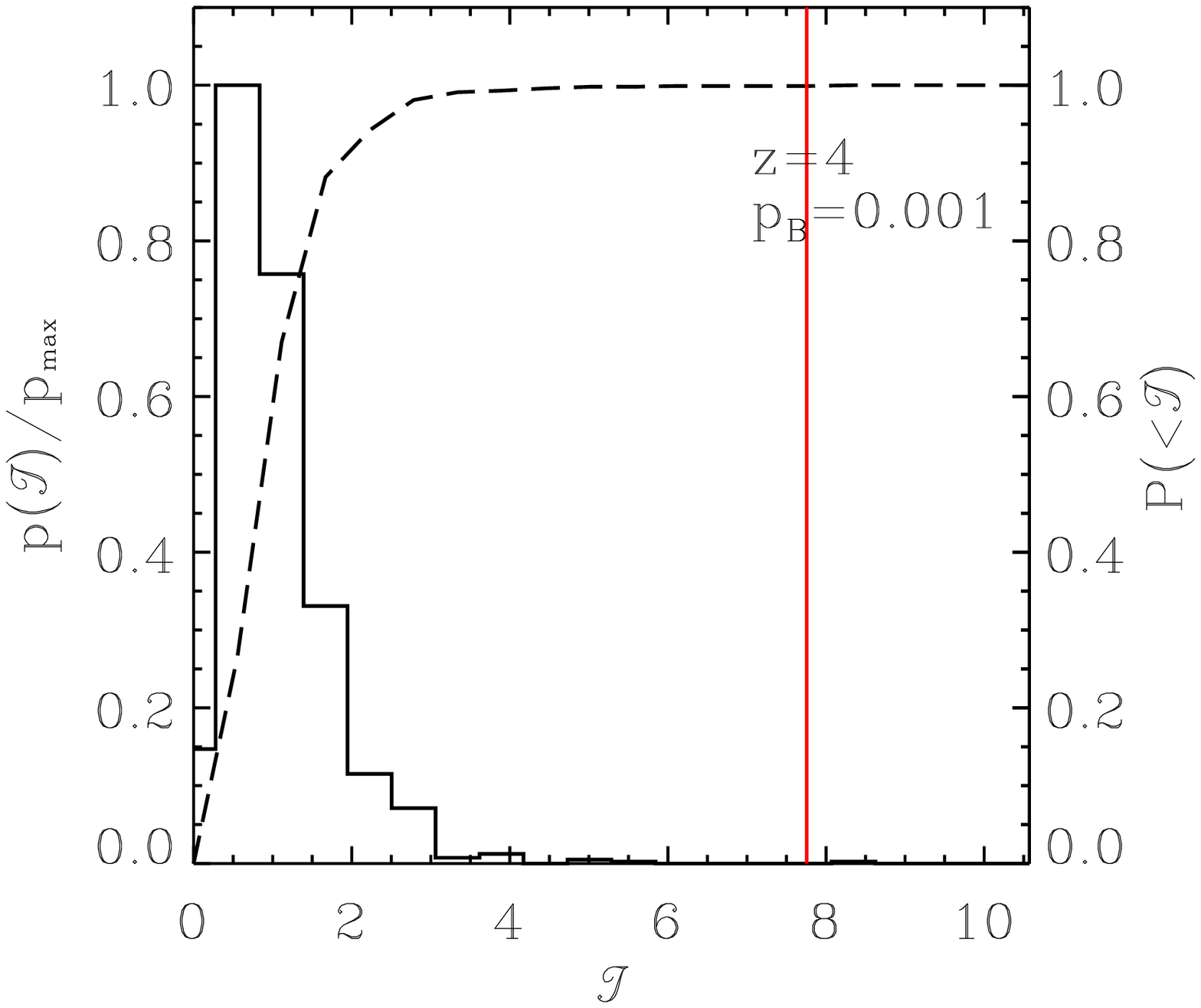, scale=0.4}
   \end{center}
 \end{minipage}
 \hfill

\caption{\large The posterior predicted distribution of the test
  quantity $\calT$ for the galaxy mass functions at four different redshifts. 
The $p$-value for each redshift is labelled in the corresponding panel.
}\label{fig:PCA_smfz}
\end{figure}

With the advent of large and deep surveys of galaxies, 
the evolution of the galaxy stellar mass function can now 
be observed to $z\sim8$ 
\citep[e.g.][]{Bouwens2009, Labbe2010, Oesch2010, Yan2010}. 
Here, we use our constrained model to predict the stellar 
mass functions at $z=0$, $1.15$, $2.5$ and $4$, 
and compare our model predictions with the existing data. 
We do not consider data at $z>4$ because they are still 
quite uncertain. For each of the four redshifts, 
we use $10^4$ halos, sampled from a mass distribution 
$({\rm d}N/{\rm d}m) \propto m^{-1.5}$, to construct 
merger trees rooted at that redshift, and adopt our posterior parameter
distribution to predict the galaxy population using those merger trees. 
We then assign a weight to each predicted 
galaxy according to the ratio between the 
halo mass function at that observed redshift [again estimated using the 
\citet{Sheth1999} formula] and the mass distribution
used to sample the merger trees. The reason for sampling the
halo merger trees in this way instead of just using the halo 
mass function is to guarantee that the massive halos 
are well sampled. The stellar mass function at a given 
redshift is then obtained through the weighted counts of 
the predicted galaxies at the redshift in question. 
Figure \ref{fig:smfz} shows the predicted stellar mass 
function at $z=0, 1.15, 2.5$ and $4$ compared 
with the observational data. The stellar mass function 
at $z=0$ is adopted from \citet{Li2009};
the $z=1.15$ mass function is that given by  
\citet{Perez-Gonzalez2008} for galaxies in the 
redshift range from $1$ to $1.3$; the $z=2.5$ mass function 
is that of \citet{Marchesini2009} for galaxies in the 
redshift range from 2 to 3; and the $z=4$ mass function 
is that of \citet{Stark2009} for galaxies with 
$3.2<z<4.7$. All the stellar mass functions are converted 
to the Chabrier IMF \citep{Chabrier2003} used in 
our model. Because our model is constrained by the $K$-band 
luminosity function at $z=0$, it is not surprising 
that the predicted stellar mass function at $z=0$ agrees 
with the observations. For higher
redshifts, the model predictions show a
larger discrepancy with the observations, although the 
posterior predictive distributions are quite broad.  
However, these broad distributions are misleading, 
because the shapes of the predicted stellar mass functions 
are systematically different from those observed, as shown 
by the predictions of 8 randomly-selected individual 
models, also plotted in Figure \ref{fig:smfz} at each redshift.   
To quantify the differences between the model
predictions and the observational results, we again use the 
PPC described in \S\ref{sec:bayes_check}. The results, 
shown in Figure \ref{fig:PCA_smfz}, clearly demonstrate 
that the differences are significant for all three high-$z$
cases.

\begin{figure}
	\begin{center}
	\epsfig{file=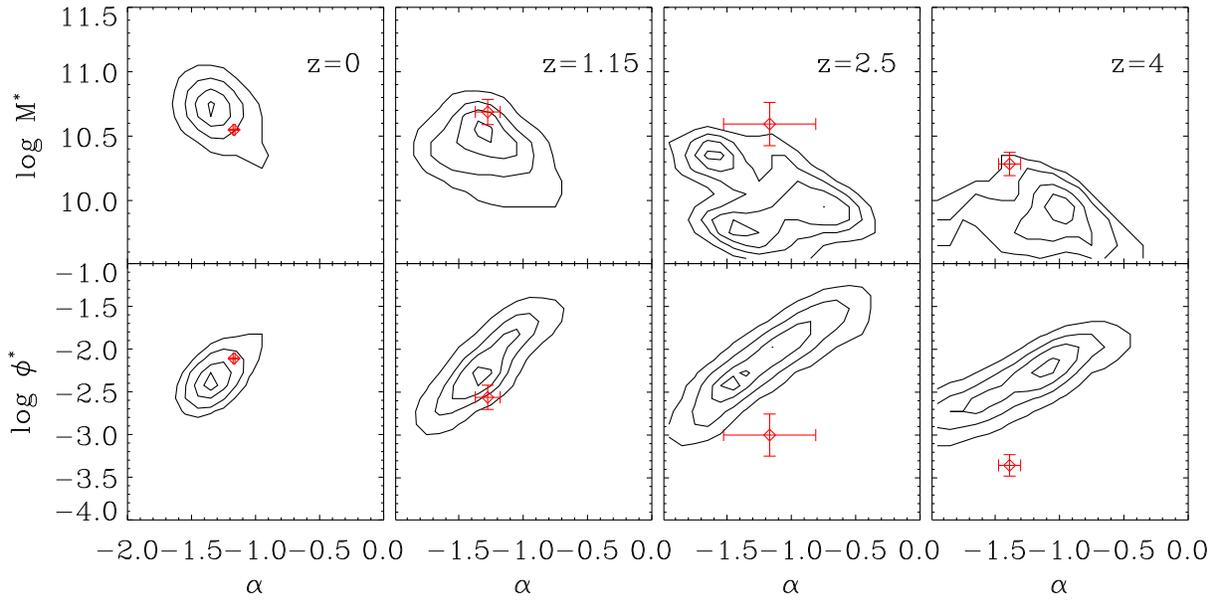, scale=0.7}
	\end{center}
\caption{\large
The posterior predictive distribution of the Schechter function 
parameters of the predicted stellar mass functions 
at $z=0, 1.15, 2.5$ and 4. The contours enclose the 5\%, 33\%, 67\% 
and 95\% confidence levels. The red crosses denote the fitted 
values for the corresponding observational results. 
}\label{fig:fit_smfz}
\end{figure}

Although the test based on the Principal Components is 
powerful and general, it is not easy to see in which aspects
the model prediction fails.  Since the stellar mass functions are 
usually  characterised by a Schechter function, it might 
be interesting to have a PPC based on the Schechter parameters.  
Figure \ref{fig:fit_smfz} shows the predictive posterior 
distributions of the Schechter parameters (contours) compared 
to the observational values (red crosses with 1-$\sigma$ 
error bars). Here $\alpha$ is the faint-end slope of the Schechter 
function, and $\log M^*$ and $\log \phi^*$ are 
the logarithms of the characteristic mass and normalisation, 
respectively. The contours, from inside out, denote the 5\%, 33\%, 
67\% and 95\% confidence levels. 
It is clear that our model predictions agree with the $z=0$ observations
very well, but deviate from the observations for high-$z$ galaxies.   
The characteristic stellar masses of our predicted galaxies are 
systematically lower than that of the observed galaxy population and 
the normalisations are systematically higher, suggesting that
the model over-predicts the number of low-mass galaxies
and under-predicts the number of high-mass galaxies.  These results are
consistent with those presented in \citet{Bower2006}, \citet{Kitzbichler2007}
and \citet{Guo2011b}. However, these authors only 
showed the predictions for individual parameter sets,  
while ours are based on the entire posterior distribution
of the model parameters.

\begin{figure}
    \begin{center}
      \epsfig{file=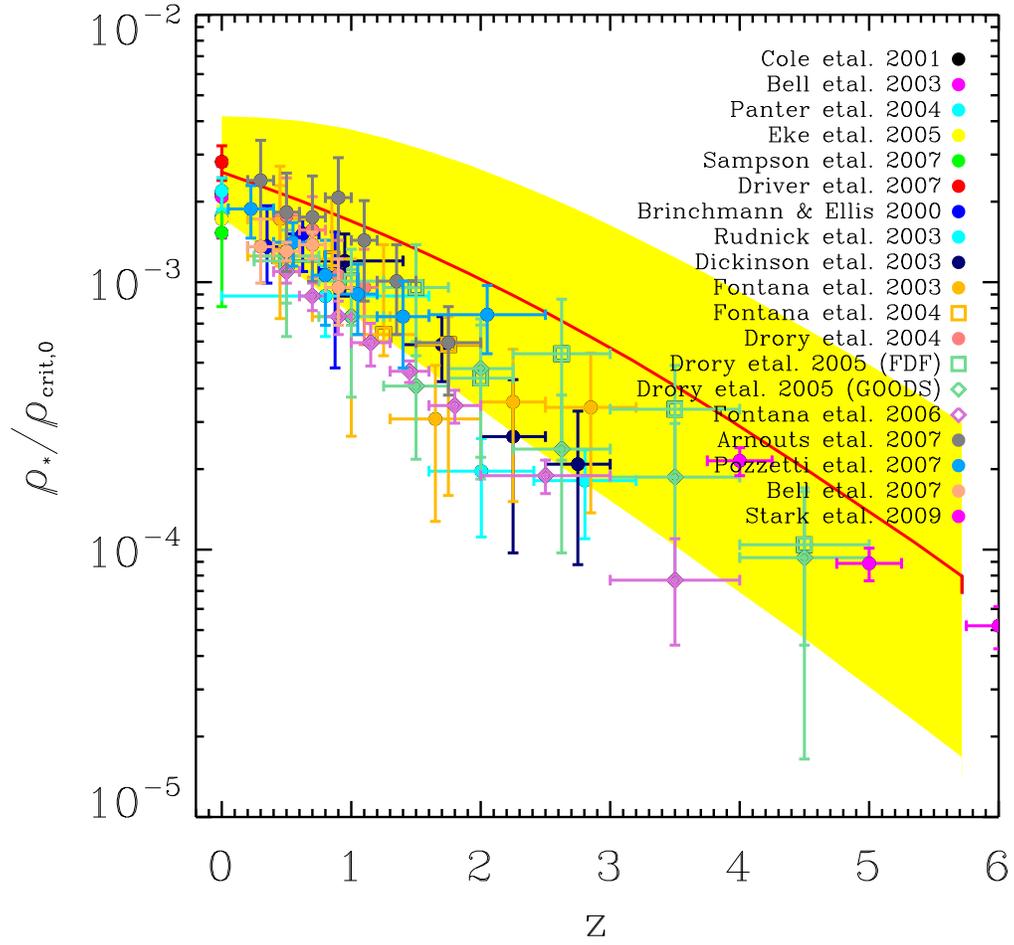, scale=0.8}
    \end{center}
\caption{\large 
The model predictions for the comoving stellar mass
density of the universe normalised by the present day critical
density compared with observational data.
The yellow band encompasses the 95\%
confidence range and the red solid line shows the median value
of the predictions. The points with error bars show various
observational estimates.
}\label{fig:smd}
\end{figure}

 We also predict the redshift evolution of the total stellar 
mass density of the universe using merger trees rooted 
at $z=0$. Since the mass resolution of the merger trees 
for all the halos is set at $1\times10^{9}\msunh$, 
the progenitor halos are well sampled at high redshifts.
At any given redshift, the stellar mass of each modelled
galaxy is weighted according to the mass of its descendant halo at $z=0$
and the \citet{Sheth1999} mass function at $z=0$.
The stellar mass density is then obtained by summing up  
the weighted stellar masses of all the modelled galaxies 
with stellar masses larger than $10^8\msun$.
Figure~\ref{fig:smd} shows the predicted comoving
stellar mass density normalised by the critical 
density of the Universe at the present time,
together with the observational results presented in
\citet{Wilkins2008} and \citet{Stark2009}.
To make a fair comparison, we correct the data points by 
taking into account the effects of using different IMFs and 
SPS models. The data in \citet{Wilkins2008} assumed a Salpeter IMF
\citep{Salpeter1955}, making the stellar masses about 70\% 
higher than those one would derive using a Chabrier IMF
\citep{Wilkins2008}. The data in \citet{Wilkins2008}
were obtained from various observational measurements
using the BC03 SPS model \citep{Bruzual2003} or 
the PEGASE  SPS model \citep{Fioc1997}, 
which includes less contributions from AGB stars than the CB07 model 
and can also over estimate the stellar mass by 5-25\%,
depending on the star formation history in the past Gyr
\citep{Conroy2010}. In our model-data comparison 
shown in Figure~\ref{fig:smd}, these two effects 
are approximately included by shifting the data 
points downwards by a factor of $1.9$. As one can see, 
the predicted stellar mass density at $z=0$ matches 
the observational results well. Again, this is not surprising, as   
the model is constrained by the $K$-band luminosity 
function at the present time. Moreover, the predicted 
stellar mass density decreases with increasing 
redshift, a trend similar to that in the data.
However, the model predictions are systematically higher 
than the observational data at $z>0$.   

\begin{figure}
    \begin{center}
      \epsfig{file=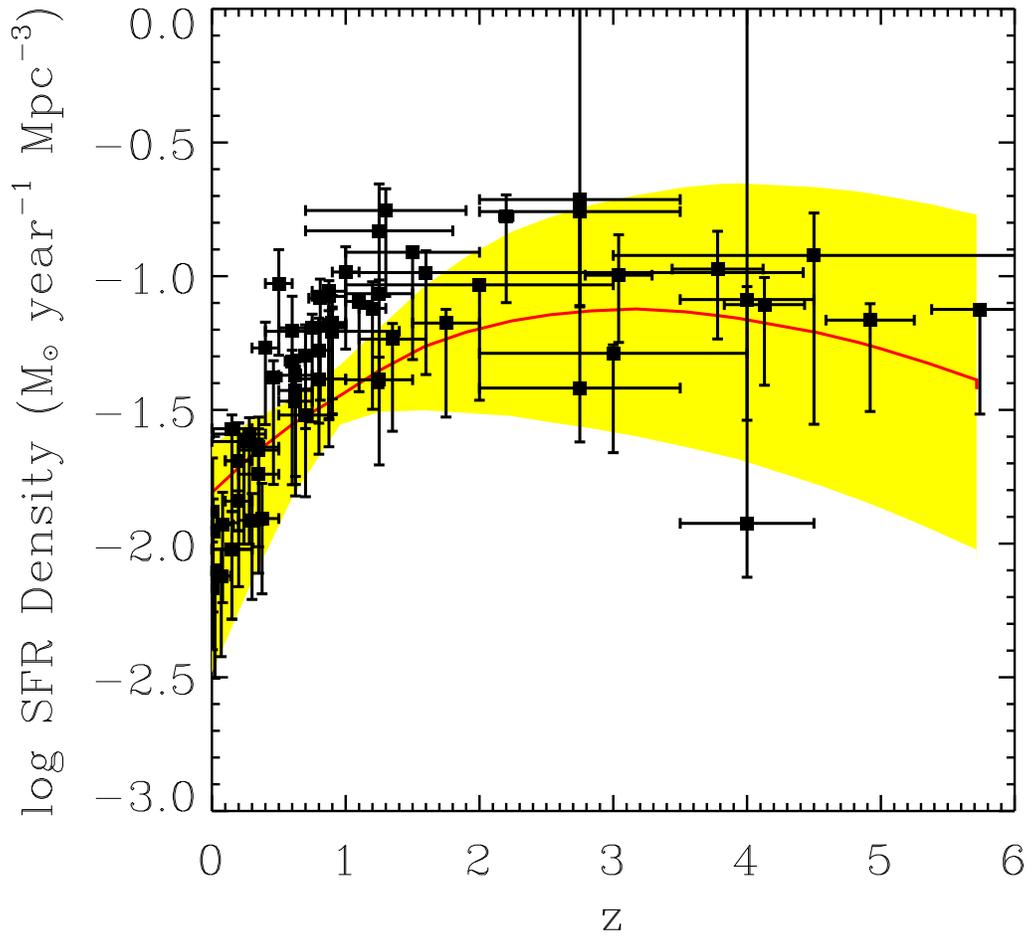, scale=0.8}
    \end{center}
\caption{\large 
The posterior predicted comoving star formation
rate density of the universe.  The yellow band encompasses the 95\%
confidence range, and the red solid line shows the median values
of the predictions. The points with error bars show
observational estimates.
}\label{fig:sfrd}
\end{figure}

In Figure \ref{fig:sfrd} we compare the model 
predictions for the star formation rate density as a 
function of redshift with the data collected in 
\citet{Hopkins2004}. Since the data in
\citet{Hopkins2004} are based on the assumption of
a Salpeter IMF, we shift the data points downwards
by a factor of 1.59 to account for differences
in the star formation rates between the Salpeter 
IMF and the Chabrier IMF \citep{Leroy2008}. 
The model predictions agree with the observational 
results at $z=0$. This is not trivial because the star 
formation rate is {\it not} used as a constraint.
It is also remarkable that the model reproduces the
overall trend of the evolution. In detail, 
the predicted increase of the star formation rate 
density with redshift below $z=1$ appears slower 
than that in the observations. At higher redshifts,
the predicted star formation rate density 
declines mildly with redshift, while the observations 
show a roughly constant rate over a large range of 
redshift. The largest discrepancy occurs in the
redshift range between $0.5$ and $3$, where most 
of the data points lie above the model predictions. 
This suggests that the model underpredicts the star 
formation rate at high redshift. However, as we 
have shown above, the same model actually 
{\it overpredicts} the stellar mass density.
Thus, the discrepancy between the model 
predictions and the data cannot be solved simply by 
changing the overall star formation or feedback 
efficiencies. It indicates that the data sets of 
the stellar mass density and the star formation rate 
density are mutually inconsistent either because 
of uncertainties in the observations or because 
some assumptions used to derive the stellar mass and 
star formation rate from observables may be incorrect.
For example, the data of the stellar mass and star 
formation rate are all derived with the assumption that 
the IMF is universal, while in reality the IMF may vary
with redshift \citep{Dave2007, Fardal2007}. 
Indeed, it has been demonstrated that if the IMF 
is more ``top-heavy'' (or ``bottom-light'') at higher redshifts, 
the discrepancy between the star formation 
rate and the stellar mass density can be alleviated \citep{Wilkins2008, Kang2010}. 
Our results demonstrate that, even if the model parameters 
are varied over a large range, the current model 
family (which assumes a universal IMF) may still
not be able to match the data over the observed redshift 
range, suggesting that a redshift-dependent IMF 
might be necessary.

\begin{figure}
\epsfig{file=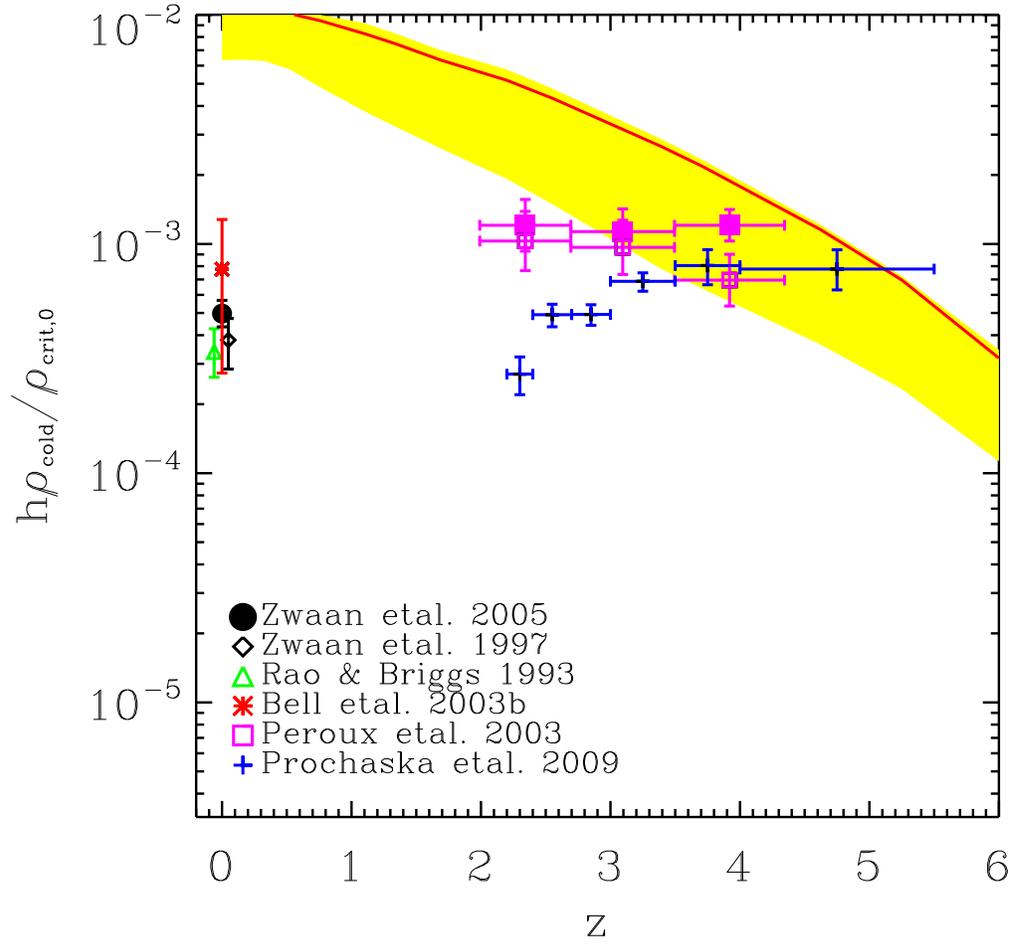,scale=0.8}
\caption{\large 
The posterior predicted comoving cold gas mass
density of the universe normalised by the critical density of
of the universe at the present time.  The yellow band encompasses the 95\%
confidence range of the model predictions while
the red solid line shows the median. Points with error bars 
show observational estimates.
}\label{fig:cgmd}
\end{figure}

In Figure \ref{fig:cgmd}, we show the predictions of the comoving 
cold gas mass density, normalised by the critical density 
of the Universe at $z=0$, as a function of redshift. Here again we make
corrections for the contributions of ${\rm He}$
and ${\rm H_2}$ using the simple models described earlier. 
We compare with observational results at $z\sim 0$ 
either from HI gas surveys of local galaxies \citep{Rao1993,Zwaan1997, Zwaan2005a} 
or from empirical models \citep{Bell2003a}, and with high-redshift 
measurements based on  DLA systems \citep{Peroux2003, Prochaska2009}. 
Once again the model significantly overpredicts the cold gas mass at  
low redshifts.  In particular, the model predicts an increasing trend 
of the cosmic cold gas mass density with decreasing redshift,  
whereas the data show that the cold gas density actually decreases 
with time at $z<2$.  This suggests that any processes that reduces
the cold gas content of galaxies must be time-dependent, operating 
effectively only at relatively low redshifts. The preheating model 
advocated by \citet[][]{Mo2005} has this property, and we will 
use our Bayesian SAM to explore this possibility in an upcoming paper.

\section{Discussion}\label{sec:discussion}

We have used a Bayesian SAM of
galaxy formation to make model inferences from the observed $K$-band
luminosity function of galaxies. We found that some of the free parameters specifying our model
family are well constrained even with this single data set,
and the posterior distribution contains the parameters adopted in some existing
SAMs. We have used the posterior distribution to make predictions for
the colour-magnitude relation of galaxies, the Tully-Fisher relation,
the conditional stellar mass function of galaxies in halos of
different masses, the HI mass function, the redshift evolution of the
stellar mass function of galaxies, and the star formation history, all
with their full inference uncertainties.  The information in the
available data can be used to check the model. Comparing the model
predictions with available observational results we have found that
the current model family, although covering a large model space, still
has serious tensions with the observational data. It
over-predicts the satellite fraction, and
vastly over-predicts the HI mass function at $z\sim 0$.  It predicts
stellar mass functions that are too steep at high redshift. It
predicts a redshift evolution of the stellar mass density and the star
formation history of the Universe that are in 
conflict with the observations.  It predicts Tully-Fisher relations
that are not well described by a pure power-law relation between
galaxy stellar mass and rotation velocity. These discrepancies suggest
that the current model family may still miss some processes important
for galaxy formation and evolution.

Current SAMs
over-predict the satellite fraction. Comparing the conditional stellar
mass functions of galaxies predicted by four popular SAMs with the
observational results of \citet{Yang2009a}, \citet{Liu2010} found that
all of the SAMs over-predict the satellite fraction by a factor of two
or more.  Since our model family covers a large parameter space, our
results demonstrate that this is a generic problem for the current
model family, rather than just for the specific models considered by
\citet{Liu2010}.  There are at least two ways to
address this problem, both requiring an extension of
the current model family to include some new processes. 
The first is to suppress star formation in dark matter halos at high
redshift. However, this will further exacerbate the current
underprediction in the star formation history at high-$z$ (see
Fig.\,\ref{fig:sfrd}).
The second is to introduce tidal stripping
and disruption to reduce the number of satellites.
Observationally, there are indications that some satellite galaxies
are being destroyed by the tidal forces of their hosts and/or by
interactions with substructures in their hosts
\citep[e.g.][]{Mihos2005}.  In addition, recent observations have
revealed the existence of halo stars in clusters and groups of
galaxies \citep{Zibetti2005, Gonzalez2005, Krick2006, Zibetti2008},
which are believed to be stars stripped from satellite galaxies.  As
discussed in \citet{Yang2009} and \citet{Liu2010}, the observed
satellite population can be better reproduced when one allows for
a halo component of stars.  In a forthcoming paper, we will use the
observed conditional luminosity functions of galaxies as additional
constraints, and use Bayesian evidence to examine whether a new model
family including tidal disruption is favoured over the original model
family. The resulting posterior will then be used to predict the
amount of halo stars in halos of different masses and to check whether
the model predictions are consistent with observations.

 The current model family constrained by the $K$-band luminosity
function vastly over-predicts the HI-mass function. This problem
has not been widely recognised \citep[but see][]{Mo2005}, as many
of the early investigations focused only on the stellar component.
In the current $\Lambda$CDM model considered here, the baryon
component is about $17\%$ of the total mass density of the
universe, while only a small fraction, $\approx10\%$, of the baryons
are in stars. In low-mass halos where most of the baryonic matter 
is accreted through the cold-mode accretion
\citep{Birnboim2003,Kerevs2005,Kerevs2009} 
and radiative cooling is very efficient,
the fraction of the gas that has not been locked into stars
must be in the cold phase, unless the gas is heated or ejected
by some feedback process. In most SAMs, including the model
family considered here, gas accretes into
dark matter halos and cools, but then can be heated in and/or ejected
from halos by supernova explosions associated with star
formation.  However, the total energy produced is
limited. As we have seen, the energy required to reduce
star formation sufficiently is already a large fraction of the
total energy produced. To remove most of the cold gas so
that the resulting HI-mass function matches the observed one
requires even higher efficiencies. Even worse, numerical
simulations have demonstrated that the efficiency of
supernova feedback in reducing the cold gas in low-mass
galaxies is actually very low \citep{MacLow1999}. All of this
suggests that supernova feedback as implemented in current SAMs
may not be responsible for suppressing star formation
in low-mass halos \citep{Mo2005}. One alternative possibility,
proposed by \citet{Mo2002, Mo2004} and \citet{Mo2005} 
but not yet thoroughly investigated, is the preheating of
the intergalactic gas, which results in a reduced fraction
of the gas that can be accreted by low-mass halos \citep{Lu2007}.
Such preheating might be produced by star formation and/or AGN
activity during early phases of rapid star formation
\citep{Mo2002, Mo2004}, or by shocks associated with
the formation of large-scale pancakes in the cosmic
density field \citep{Mo2005}. In a forthcoming paper,
we will explore such a model family.

The current model family also predicts stellar mass
functions for high-$z$ galaxies that is different than those observed,
i.e. it predicts too many low-mass galaxies and an insufficient number of
massive galaxies.  There are at least two explanations for
this discrepancy. First, the current model family might not properly
take into account the redshift dependence of star formation.  In our
model, merger-driven star bursts are distinguished from quiescent star
formation, so that a redshift-dependence of star formation owing to
the redshift dependence of the galaxy merger rate naturally occurs.
However, it appears that this effect alone is
insufficient.  Another process that may
lead to a redshift-dependent star formation rate is the accretion of
cold gas into galaxies.  
Gas accretion at high-$z$ is dominated by cold-mode accretion, while
hot-mode accretion dominates at low-$z$
\citep{Kerevs2005,Kerevs2009}. Gas accretion, and hence star
formation, proceeds faster in galaxies with cold-mode accretion than
in those with hot-mode accretion, so more stars would form at high
$z$.  Our current model family does not distinguish between cold and
hot mode accretion explicitly, and it would certainly be interesting
to explore models that do.  Second, the merger time scales at high-$z$ 
may be overestimated relative to those at low-$z$. If the mergers of
low-mass galaxies occurred more frequently at high $z$, the number of
small galaxies might be reduced, while the number of massive galaxies
would be increased by the larger number of merger remnants. 
It is possible that the Chandrasekhar dynamical friction formula 
overestimates the merger time scales because accretion 
at high-$z$ is dominated by mergers along a few filaments,
making the galaxies embedded in them merge with the central galaxy
faster. It would be interesting to quantify such effects with
numerical simulations and to include them in a semi-analytic model to
explore its impact on the redshift evolution of the galaxy luminosity
function.
 
The current model family predicts a star formation history that is
lower than that observed at high $z$.  Since our model predictions
match the stellar density at the present epoch, simply increasing the
star formation rate at high-$z$ would over-predict the total stellar
mass density at the present time. There is also not much room to
reduce the star formation rate at low-$z$ to compensate for the
increase at high-$z$, as the current model predicts a star formation
history at low-$z$ that matches the observations.
A redshift-dependent stellar initial mass function (IMF) may
address this discrepancy.  If the IMF is top heavy (or bottom light) at high
redshift because of a different star formation mode,
e.g. in merger driven starbursts, the observed star formation history
for a universal IMF would
overestimate the true star formation rate.  This would help
alleviate the discrepancy between our model predictions and
observations. However, a non-universal IMF would have other
observational consequences and a systematic analysis of such
consequences would be required to show that such an IMF is indeed
preferred.

Finally the current model family predicts Tully-Fisher relations
that are curved, suggesting that either selection effects 
in the observational samples are not properly taken into
account in the model or an interaction between the baryonic and dark matter 
components plays a crucial role in shaping the observed 
Tully-Fisher relation. 
It is important, as a next step, to include a detailed model for 
the rotation velocities, 
and to compare the model predictions with an observational sample
where the selection effects are better understood \citep[e.g][]{Pizagno2007}.

Our Bayesian SAM allows us to explore the various possibilities 
mentioned above with probabilistic rigour. The Bayesian
`goodness of fit' provided by the posterior
predictive check helps assess the admissibility of
model families, and Bayesian evidence can be used to discriminate
between different model families using the observational data.  In a
series of forthcoming papers we will use these tools of Bayesian
inference to address the various problems identified above.


\section*{Acknowledgement}
We thank Aaron Dutton for providing the Tully-Fisher relation data, 
Xiaohu Yang for providing the SDSS DR7 catalogue with 
magnitudes and colours, and Peter Behroozi for providing the stellar mass functions. 
We also thank David Weinberg, Michel Fioc, Stephane Charlot, Zhankui Lu and Xi Kang for useful comments. 
This material is based upon work supported by NASA grant AISR-126270,
NSF grant IIS-0611948, and NSF grant AST-1109354.

\bibliography{/Users/luyu/references/general}

\newpage
\begin{table}
\caption{Model parameters} 
\centering
\begin{tabular}{l c c c c}
\hline\hline
\# & Parameter & Meaning & Prior 
\\ [0.5ex]
\hline
1 & $\log M_{\rm CC} (\msun) $ & cooling cut-off halo mass& [1.5 , 4.5] \\
\hline
2 & $\log \alpha_{\rm SF}$ & star formation efficiency power-law amplitude & [-3, 0] \\
\hline
3 & $\beta_{\rm SF}$ & star formation efficiency power-law index &  [-1, 12] \\
\hline
4 & $\log V_{\rm SF} $ (km/s) & star formation law turn-over halo circular velocity & [1.5, 3.0] \\
\hline
5 & $\log f_{\rm SF} (\msun/{\rm pc}^2)$ & star formation threshold gas surface density & [-1, 3] \\
\hline
6 & $\log \alpha_{\rm SN}$ & SN feedback energy fraction & [-3, 1] \\
\hline
7 & $\log \alpha_{\rm RH}$ & SN feedback reheating power-law amplitude & [-3, 2] \\
\hline
8 & $\beta_{\rm RH}$ & SN feedback reheating power-law index & [0, 14] \\
\hline
9 & $\log \epsilon_{\rm W}$ & fraction of surplus SN feedback energy used for powering wind & [-3, 0] \\
\hline
10 & $\log f_{\rm RI}$ & fraction of re-infall ejected hot gas & [-2, 0] \\
\hline
11 & $\log f_{\rm DF}$ & merging time-scale in dynamical friction time-scale & [0, 2] \\
\hline
12 & $\log \alpha_{\rm SB}$ & merger triggered star burst efficiency power-law amplitude& [-2, 0] \\
\hline
13 & $\beta_{\rm SB}$ & merger triggered star burst efficiency power-law index & [0, 2] \\
\hline
14 & $\arctan(\alpha_{\rm IN})$ & faint-end incompleteness & [0, 0.177] \\
\hline
\hline 
\end{tabular}
\label{tab:param}
\end{table}


\newpage

\newpage


\newpage

\newpage

\newpage


\newpage


\newpage

\newpage

\newpage


\newpage


\newpage


\newpage

\label{lastpage}

\end{document}